\documentclass[nofootinbib]{revtex4-2}
\usepackage{bm}
\usepackage{amsmath}
\usepackage{amssymb}
\usepackage{graphicx}
\usepackage[caption=false]{subfig}
\usepackage{todonotes}
\usepackage{mathtools}
\usepackage[breaklinks,colorlinks]{hyperref}
\usepackage{placeins}

\hypersetup{%
,urlcolor=blue
,citecolor=blue
,linkcolor=blue
}

\numberwithin{equation}{section}

\begin{document}

\title{Pinching instabilities in superconducting cosmic strings}

\author{R.~A.~Battye}
\email[]{richard.battye@manchester.ac.uk}
\affiliation{%
Jodrell Bank Centre for Astrophysics, School of Natural Sciences, Department of Physics and Astronomy, University of Manchester, Manchester, M13 9PL, U.K.
}

\author{S.~J.~Cotterill}
\email[]{steven.cotterill@postgrad.manchester.ac.uk}
\affiliation{%
Jodrell Bank Centre for Astrophysics, School of Natural Sciences, Department of Physics and Astronomy, University of Manchester, Manchester, M13 9PL, U.K.
}

\label{firstpage}

\date{\today}

\begin{abstract}
Superconducting cosmic strings can exhibit longitudinal, pinching instabilities in some regions of the parameter space. We make predictions about the onset of this instability using the thin string approximation (TSA) and develop an improved analysis that remains applicable for small wavelength perturbations, where the TSA breaks down. We use simulations of perturbed strings to assess the accuracy of the TSA, test the predictions of our new analysis and demonstrate an improvement over previous methods in the literature. Notably, it appears that the instabilities are typically present for a larger range of magnetic strings than previously expected, and we show examples of pinching instabilities also occurring in electric strings. However, both our simulations and predictions agree that strings near the chiral limit are free from pinching instabilities and in particular our results support our previously published claim that vortons can be stable to all classical perturbations if they are sufficiently large.
\end{abstract}

\maketitle

\section{Introduction}

Cosmic strings are a well studied, one-dimensional topological defect that often appear in models that attempt to go beyond the Standard Model of Particle Physics, see \cite{V&Sbook,Hindmarsh_1995,Vachaspati:2015} for reviews. A path in space that encircles a string will loop around the vacuum manifold $n$ times, where $n$ in known as the winding number. Witten showed \cite{Witten1985} that strings can host superconducting currents in their cores (and it has since been argued that strings are generically current carrying \cite{DAVIS1995197}) which can cause high energy, potentially observable events, for example see \cite{PhysRevLett.101.141301}, and have interesting cosmological consequences. In particular, the formation of static loops supported by angular momentum - known as vortons \cite{Davis1988b,Davis1988c} - that can, depending on the specific model, either overclose the Universe or contribute to dark matter \cite{Brandenberger1996,Martins1998b,Auclair:2020wse}.

Under certain conditions, the current on the strings can become unstable, leading to potentially observable radiation and the likely destruction of vortons. This effect was discussed in \cite{Lemperiere2003a} where an analytical criterion was derived for the onset of this pinching instability for magnetic strings, that was based on perturbations to field theory solutions. In an alternative approach, which we refer to as the thin string approximation (TSA), the string is treated as having negligible width and there can be a longitudinal instability that occurs when the square of the longitudinal sound speed becomes negative \cite{Carter1989,Carter1993}. Dynamical simulations have been performed using this formalism and instabilities were found in the magnetic regime that manifested themselves as shocks in the equation of state parameter \cite{PhysRevD.61.043510}.

We recently presented evidence of a vorton that was fully stable based on predictions from the TSA and supporting numerical simulations \cite{PhysRevLett.127.241601,2022JHEP...04..005B}. However, small radii vortons that were predicted to be stable suffered from an unexpected pinching instability that disappeared as the radius of the vorton was increased. In this paper, we strengthen our claim that this instability is due to curvature effects (and will therefore not be present in larger vortons once the effects of curvature become negligible) by performing simulations of straight superconducting strings in the same parameter set and showing that they are stable near the chiral limit.

Notably, both of the previous methods for predicting the onset of pinching instabilities integrate over the width of the string, removing degrees of freedom from the system. As such, one should expect these methods to break down when the width of the string can no longer be treated as negligible. We have developed an improved analysis for dealing with these cases and assess the accuracy of all three methods when compared with simulations.

We will be using the neutral limit of a gauged $U(1)\times U(1)$ model with the Lagrangian density,
\begin{equation} \label{eq: Lagrangian}
    \mathcal{L} = (\mathcal{D}_\mu\phi)(\mathcal{D}^\mu\phi)^* + \partial_\mu\sigma\partial^\mu\sigma^* - \frac{1}{4}F_{\mu\nu}F^{\mu\nu} - \frac{\lambda_\phi}{4}(|\phi|^2 - \eta_\phi^2)^2 - \frac{\lambda_\sigma}{4}(|\sigma|^2 - \eta_\sigma^2)^2 - \beta|\phi|^2|\sigma|^2,
\end{equation}
where $\mathcal{D}_\mu = \partial_\mu - igA_\mu$ and $F_{\mu\nu} = \partial_\mu A_\nu - \partial_\nu A_\mu$. The parameters are all real positive constants and are chosen so that $\text{U(1)}_\phi$ is broken in the vacuum and $\text{U(1)}_\sigma$ is only broken along the core of the string, where it condenses. We will always use $\eta_\phi = \lambda_\phi = 1$ since all other systems can be obtained simply by rescaling the other parameters and the length scales - see \cite{2022JHEP...04..005B} for more information. We also parameterise the gauge coupling with $G = g/g_\text{BPS}$ where $g_\text{BPS}^2 = \lambda_\phi/2$ is the coupling required to set the length scales of the vortex field and gauge field to be equal. For a string lying along the $z$-axis, we can make the ansatz that $\phi = |\phi|(\rho)\exp[in\theta]$ (although we typically use $n=1$), $\sigma = |\sigma|(\rho)\exp[i(\omega t + kz)]$ and $A_\theta = A_\theta(\rho)$, with all other gauge field components being zero everywhere, for which static solutions satisfy,
\begin{align}
    \frac{d^2|\phi|}{d\rho^2} + \frac{1}{\rho}\frac{d|\phi|}{d\rho} - \bigg[\frac{1}{2}\lambda_\phi(|\phi|^2 - \eta_\phi^2) + \beta|\sigma|^2 + \bigg(\frac{n-gA_\theta}{\rho}\bigg)^2\bigg]|\phi| &= 0 \label{eq: 1D EoMs start}, \\
    \frac{d^2|\sigma|}{d\rho^2} + \frac{1}{\rho}\frac{d|\sigma|}{d\rho} - \bigg[\frac{1}{2}\lambda_\sigma(|\sigma|^2 - \eta_\sigma^2) + \beta|\phi|^2 - \chi \bigg]|\sigma| &= 0 \label{eq: sigma straight static EoM}, \\
    \frac{d^2A_\theta}{d\rho^2} - \frac{1}{\rho}\frac{dA_\theta}{d\rho} + 2g|\phi|^2(n - gA_\theta) &= 0 \label{eq: 1D EoMs end},
\end{align}
with $\chi = \omega^2 - k^2$. Strings are categorized based on the sign of $\chi$ - electric strings have $\chi>0$, chiral strings have $\chi = 0$ and magnetic strings have $\chi<0$. Due to the Lorentz invariance of the string under boosts in the $z$ direction, one can shift to a frame in which $\omega$ = 0 for magnetic strings, or one where $k=0$ for electric strings, while $\chi$ remains the same. We will call these frames the pure magnetic or pure electric frames respectively. There are two conserved quantities that will be of interest - the Noether charge associated with the $\text{U(1)}_\sigma$ symmetry, $Q$, and the topological winding number of the condensate along the string, $N$, defined as
\begin{equation}
    Q = \frac{1}{2i}\int d^3x (\sigma^*\partial_t\sigma - \sigma\partial_t\sigma^*) = 2\pi\omega L\int|\sigma|^2\rho d\rho  \qquad \text{and} \qquad N = \frac{kL}{2\pi},
\end{equation}
where $L$ is the length of string.

\begin{figure}[!t]
    \centering
    \subfloat[$\eta_\sigma=0.1825$, $\lambda_\sigma=900$, $\beta=20$ and $G=0.2$ (denoted parameter set G in \cite{2022JHEP...04..005B}) with $\chi=-3$.]{
        \centering
        \includegraphics[trim={0.5cm 0 1.5cm 0},clip,width=0.46\linewidth]{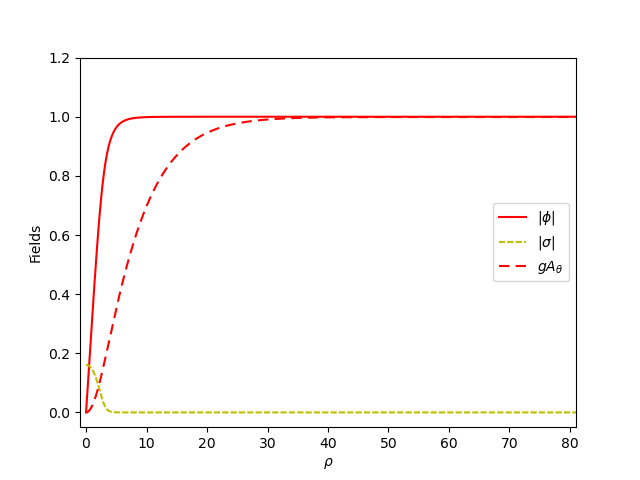}
        \label{fig: pG straight string profile}
    }\hspace{1em}
    \subfloat[$\eta_\sigma=1$, $\lambda_\sigma=2/3$, $\beta=2/3$ and $G=0.1$ (denoted parameter set E in \cite{2022JHEP...04..005B}) with charge per unit length in the purely electric frame, $q_p=33.4$ ($\chi\approx 0.135$).]{
        \centering
        \includegraphics[trim={0.5cm 0 1.5cm 0},clip,width=0.46\linewidth]{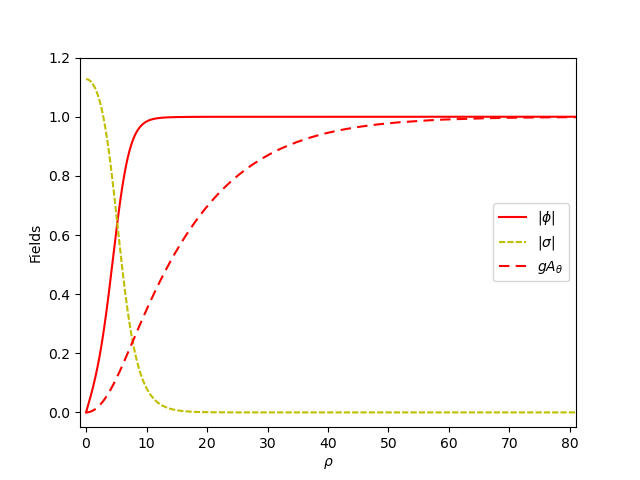}
        \label{fig: pE straight string profile}
    }
    \caption{Straight string profiles for two different parameter sets. Figure (a) is an example of a magnetic string and only has a small amplitude condensate. In contrast, figure (b) is an electric string with a large amplitude condensate that has the effect of significantly widening the core of the string. Additionally, the gauge field has double the length scale compared to figure (a), due to the reduction in the gauge coupling.}
    \label{fig:straight string profiles}
\end{figure}

One can solve these equations numerically for different choices of $\chi$, although not all values will produce a condensate on the string, see \cite{V&Sbook,2022JHEP...04..005B} for detailed discussions. Some values of $\chi$ in the electric regime can produce two solutions and it is necessary to fix the charge per unit length, $q=Q/L$, rather then $\chi$, to find these solutions. This is because one of the solutions will have lower energy than the other, which is preferred by the energy minimisation algorithms used to find the solutions, and methods based on fixed $\chi$ can reduce the charge on the string in an unphysical manner to access this lower energy state. We refer to these two solutions as either being on the lower or higher-charge branch with the expectation from the TSA being that lower-charge branch strings will be stable to pinching instabilities and higher-charge strings will not - see \cite{2022JHEP...04..005B} for a more detailed discussion. Figure \ref{fig:straight string profiles} shows a couple of examples of string solutions, that we will later show to be unstable to pinching instabilities, one of which is a magnetic string and the other is an electric string on the higher charge branch. For electric strings, one can always transform into the frame with $k=0$ and we make the choice to do this for our higher-charge branch string solutions so that they can be distinguished by the charge per unit length in this particular frame, $q_p$.

We can use the TSA to predict whether a string will be unstable to longitudinal instabilities from the equation of state of the string. We have previously shown that this method gives very accurate results when applied to the extrinsic instabilities\footnote{Extrinsic instabilities are growing oscillations in the position of the string, as opposed to intrinsic instabilities which are growing oscillations in the internal properties of the string - for example the amplitude of the condensate.} of vortons \cite{PhysRevLett.127.241601,2022JHEP...04..005B}. For the Lagrangian of equation (\ref{eq: Lagrangian}), the energy momentum tensor is,

\begin{equation} \label{eq: energy momentum tensor}
    \mathcal{T}^{\mu\nu} = 2(\mathcal{D}^\mu\phi)(\mathcal{D}^\nu\phi)^* + 2\partial^\mu\sigma\partial^\nu\sigma^* - F^\mu_{\;\alpha} F^{\nu\alpha} - g^{\mu\nu}\mathcal{L} .
\end{equation}

\noindent We can neglect the width of the string and calculate the macroscopic energy-momentum tensor, $T^{ab}$, with $a,b \in t,z$, by integrating over the string cross-section. This results in the four components,
\begin{equation}
    T^{tt} = 2\omega^2\Sigma_2 + \mu - \frac{1}{4}\lambda_\sigma\Sigma_4 , \qquad T^{tz} = T^{zt} = 2k\omega\Sigma_2 , \qquad T^{zz} = 2k^2\Sigma_2 - \mu + \frac{1}{4}\lambda_\sigma\Sigma_4 ,
\end{equation}
where we have defined the integrated quantities,
\begin{equation}
    \mu = 2\pi\int\rho d\rho \bigg\{\bigg|\frac{\partial\phi}{\partial\rho}\bigg|^2 + \bigg(\frac{n - gA_\theta}{\rho}\bigg)^2|\phi|^2 + \frac{1}{2\rho^2}\bigg(\frac{dA_\theta}{d\rho}\bigg)^2 + \frac{1}{4}\lambda_\phi(|\phi|^2 - \eta_\phi^2)^2\bigg\} \quad\text{and}\quad \Sigma_n = 2\pi\int\rho|\sigma|^nd\rho,
\end{equation}
which can be calculated for any field profiles that are a solution to equations (\ref{eq: 1D EoMs start} - \ref{eq: 1D EoMs end}). The macroscopic energy-momentum tensor can be made diagonal by transforming to the purely magnetic or electric frame, which is achieved by a Lorentz boost of velocity $v=\omega/k$ in the magnetic regime ($\omega$ is set to zero) or $v=k/\omega$ in the electric regime ($k$ is set to zero). In this frame, the tension and energy per unit length are given by $T = -T^{zz}$ and $U = T^{tt}$ respectively, and therefore
\begin{equation}
    T = 
    \begin{cases}
        \mu - \frac{1}{4}\lambda_\sigma\Sigma_4 &\mbox{if } \chi>0, \\
        2\chi\Sigma_2 + \mu - \frac{1}{4}\lambda_\sigma\Sigma_4 &\mbox{if } \chi < 0,
    \end{cases} \qquad\quad
    U =
    \begin{cases}
        2\chi\Sigma_2 + \mu - \frac{1}{4}\lambda_\sigma\Sigma_4 &\mbox{if } \chi>0, \\
        \mu - \frac{1}{4}\lambda_\sigma\Sigma_4 &\mbox{if } \chi<0.
    \end{cases}
\end{equation}
The characteristic sound speeds of transverse and longitudinal perturbations are given by $c_T^2 = \frac{T}{U}$ and $c_L^2 = -\frac{dT}{dU}$ respectively so that
\begin{equation}
    c_T^2 = \bigg(1 + \frac{2\chi\Sigma_2}{\mu - \frac{1}{4}\lambda_\sigma\Sigma_4}\bigg)^{-\text{sgn}(\chi)}, \qquad
    c_L^2 = \bigg(1 + \frac{2\chi\Sigma_2'(\chi)}{\Sigma_2}\bigg)^{-\text{sgn}(\chi)}.
\end{equation}
The conditions $c_T^2,c_L^2 \leq 1$ are necessary for causality. For all superconducting string solutions that we have found, and in the literature, the longitudinal sound speeds are smaller than or equal to the tranverse sound speeds - which is the opposite of every day, non-relativistic elastic strings and leads to more complicated dynamics for loops of string \cite{Carter1989}. 

Longitudinal perturbations with wavenumber, $p$, propagate along a straight string with frequency $\nu = c_Lp$ (and similarly for transverse perturbations). Thus if $c_L^2<0$, $\nu$ is imaginary and the perturbation is unstable. In particular, the growth rate of the instability grows linearly with the wavenumber, resulting in extreme instabilities in the short-wavelength limit. However, it must be remembered that this prediction comes from an analysis that neglects the width of the string and should be expected to break down for short wavelengths where the width is no longer negligible compared to the wavelength of the perturbation. Instead, we should expect a linear relationship between the wavenumber and growth rate at small $p$ (large wavelengths) that departs from linearity as $p$ grows larger. We will claim in this paper that these instabilities are the same as the pinching instabilities that we will observe in simulations.

\section{Stability analysis} \label{sec: gen analysis}

In principle, a more accurate analysis that takes into account the width of the string, can be performed by considering perturbations to the equations of motion, although it will be much more complicated to determine whether a particular string is unstable than using the TSA. The ansatz used to derive the static equations (\ref{eq: 1D EoMs start} - \ref{eq: 1D EoMs end}) is overly restrictive if we wish to understand the stability of the string to $z$ dependent perturbations. In appendix \ref{sec: ansatz} we will show that we can use the self-consistent ansatz $\phi = \phi_1(t,\rho,z)e^{in\theta}$, $\sigma = [\sigma_1(t,\rho,z) + i\sigma_2(t,\rho,z)]e^{i(\omega t + kz)}$ and $A_\theta = A_\theta(t,\rho,z)$ with the equations of motion
\begin{align}
    \frac{\partial^2\phi_1}{\partial t^2} - \frac{\partial^2\phi_1}{\partial\rho^2} - \frac{1}{\rho}\frac{\partial\phi_1}{\partial\rho} - \frac{\partial^2\phi_1}{\partial z^2} + \bigg[\frac{1}{2}\lambda_\phi(\phi_1^2 - \eta_\phi^2) + \beta(\sigma_1^2 + \sigma_2^2) + \bigg(\frac{n - gA_\theta}{\rho}\bigg)^2\bigg]\phi_1 &= 0 \label{eq: phi1 eq}, \\
    \frac{\partial^2\sigma_1}{\partial t^2} - \frac{\partial^2\sigma_1}{\partial\rho^2} - \frac{1}{\rho}\frac{\partial\sigma_1}{\partial\rho} - \frac{\partial^2\sigma_1}{\partial z^2} - 2\omega\frac{\partial\sigma_2}{\partial t} + 2k\frac{\partial\sigma_2}{\partial z} + \bigg[\frac{1}{2}\lambda_\sigma(\sigma_1^2 + \sigma_2^2 - \eta_\sigma^2) + \beta\phi_1^2 - \chi\bigg]\sigma_1 &= 0, \label{eq: sigma1 eq}\\
    \frac{\partial^2\sigma_2}{\partial t^2} - \frac{\partial^2\sigma_2}{\partial\rho^2} - \frac{1}{\rho}\frac{\partial\sigma_2}{\partial\rho} - \frac{\partial^2\sigma_2}{\partial z^2} + 2\omega\frac{\partial\sigma_1}{\partial t} - 2k\frac{\partial\sigma_1}{\partial z} + \bigg[\frac{1}{2}\lambda_\sigma(\sigma_1^2 + \sigma_2^2 - \eta_\sigma^2) + \beta\phi_1^2 - \chi\bigg]\sigma_2 &= 0, \label{eq: sigma2 eq}\\
    \frac{\partial^2 A_\theta}{\partial t^2} - \frac{\partial^2A_\theta}{\partial\rho^2} + \frac{1}{\rho}\frac{\partial A_\theta}{\partial\rho} - \frac{\partial^2A_\theta}{\partial z^2} - 2g(n-gA_\theta)\phi_1^2 &= 0, \label{eq: Atheta eq}
\end{align}
in the temporal gauge ($A_t=0$) and where $\phi_1$, $\sigma_1$, $\sigma_2$ and $A_\theta$ are all real.

We can now perturb all of the fields around their respective straight string solutions, $\phi_1 \to |\phi| + \delta\phi$, $\sigma_1 \to |\sigma| + \delta\sigma_1$, $\sigma_2 \to \delta\sigma_2$ and $A_\theta \to A_\theta + \delta A_\theta$, where it should be understood that $|\phi|$, $|\sigma|$ and $A_\theta$ now refer to solutions of equations (\ref{eq: 1D EoMs start} - \ref{eq: 1D EoMs end}) and terms beyond first order in the perturbed equations of motion have been ignored. Then by writing each perturbation in terms of its Fourier transform, $\delta\phi(t,\rho,z) = \int\delta\hat{\phi}(\rho,\nu,p)e^{i(\nu t + pz)}d\nu dp$, and using the fact that the perturbations are all real, the equations simplify into an eigenvalue problem with each mode being independent,
\begin{align}
    -\frac{\partial^2\delta\hat{\phi}}{\partial\rho^2} - \frac{1}{\rho}\frac{\partial\delta\hat{\phi}}{\partial\rho} + \bigg[\frac{1}{2}\lambda_\phi(3|\phi|^2 - \eta_\phi^2) + \beta|\sigma|^2 + \bigg(\frac{n-gA_\theta}{\rho}\bigg)^2\bigg]\delta\hat{\phi} + 2\beta|\phi||\sigma|\delta\hat{\sigma}_1 - \frac{2g}{\rho^2}(n-gA_\theta)|\phi|\delta\hat{A}_\theta &= \Lambda \delta\hat{\phi} , \label{eq: delta phi equation} \\
    -\frac{\partial^2\delta\hat{\sigma}_1}{\partial\rho^2} - \frac{1}{\rho}\frac{\partial\delta\hat{\sigma}_1}{\partial\rho} + \bigg[\frac{1}{2}\lambda_\sigma(3|\sigma|^2 - \eta_\sigma^2) + \beta|\phi|^2 - \chi\bigg]\delta\hat{\sigma}_1 + 2\beta|\phi||\sigma|\delta\hat{\phi} - 2i\xi \delta\hat{\sigma}_2 &= \Lambda \delta\hat{\sigma}_1 , \\
    -\frac{\partial^2\delta\hat{\sigma}_2}{\partial\rho^2} - \frac{1}{\rho}\frac{\partial\delta\hat{\sigma}_2}{\partial\rho} + \bigg[\frac{1}{2}\lambda_\sigma(|\sigma|^2 - \eta_\sigma^2) + \beta|\phi|^2 - \chi\bigg]\delta\hat{\sigma}_2 + 2i\xi\delta\hat{\sigma}_1 &= \Lambda \delta\hat{\sigma}_2 , \\
    -\frac{\partial^2\delta\hat{A}_\theta}{\partial\rho^2} + \frac{1}{\rho}\frac{\partial\delta\hat{A}_\theta}{\partial\rho} + 2g^2|\phi|^2\delta\hat{A}_\theta - 4g|\phi|(n-gA_\theta)\delta\hat{\phi} &= \Lambda \delta\hat{A}_\theta , \label{eq: delta A equation}
\end{align}
where we have defined $\xi = \omega\nu-kp$ and, if a given mode is a solution, the eigenvalue must be $\Lambda = \nu^2 - p^2$. Note that the Fourier transforms of the perturbation variables are now complex functions in general and so is the frequency, $\nu = \nu_1+i\nu_2$, and wavenumber, $p=p_1+ip_2$, with the real parts corresponding to oscillations and the imaginary parts causing exponential growth or decay. However, only solutions with $\nu_2^2-p_2^2>0$ are physically realistic instabilities because there will be a frame in which $p_2=0$; solutions with $\nu_2^2-p_2^2<0$ are perfectly valid from a mathematical point of view, but as there is no reference frame in which $p_2=0$, there is no physical way to excite these solutions and therefore they need not be considered.

Equations (\ref{eq: delta phi equation} - \ref{eq: delta A equation}) can be solved numerically, if the straight string solutions have been computed and a value is chosen for $\xi$, by discretising the radial direction with $n_\rho$ points and constructing a matrix which approximates the differential equations using finite differences - with some rows and columns extracted by enforcing boundary conditions. The eigenvalues and eigenvectors of that matrix may then be calculated. The eigenvector will have $4n_\rho$ components, with the first $n_\rho$ values giving an approximation to $\delta\hat{\phi}$ and similarly for the other variables. After calculating an eigenvalue, the frame dependent values of $\nu$ and $p$ can be determined by solving the simultaneous equations created from the definition of $\xi$ and setting $\Lambda = \nu^2-p^2$, which are
\begin{equation}
    \nu = \frac{1}{\chi}\left[\omega\xi\pm k\sqrt{\xi^2-\chi\Lambda}\right], \qquad p = \frac{1}{k}\left(\omega\nu-\xi\right). \label{eq: nu and p}
\end{equation}

\begin{figure}[!t]
    \centering
    \subfloat[An eigenmode of the perturbed string shown in Figure \ref{fig: pG straight string profile} with $\xi=-1.5$. This is a solution if $\nu=-0.597i$ and $p=\frac{\sqrt{3}}{2}$ in the purely magnetic frame.]{
        \centering
        \includegraphics[trim={0.5cm 0 1.5cm 0},clip,width=0.46\linewidth]{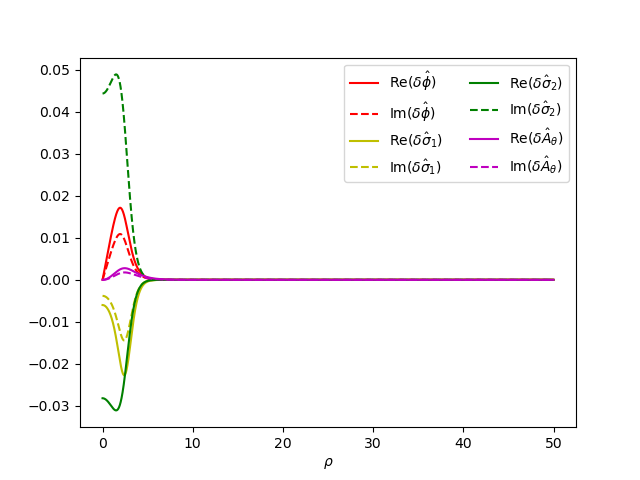}
        \label{fig: pG eigenvector}
    }\hspace{1em}
    \subfloat[An eigenmode of the perturbed string shown in Figure \ref{fig: pE straight string profile} with $\xi=0$. This is a solution if $\nu=0$ and $p=0.113$ in the purely electric frame.]{
        \centering
        \includegraphics[trim={0.5cm 0 1.5cm 0},clip,width=0.46\linewidth]{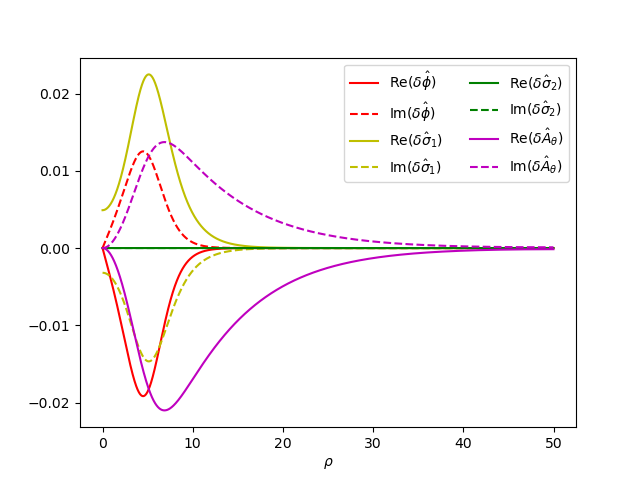}
        \label{fig: pE eigenvector}
    }
    \caption{Example eigenmodes of the strings shown in Figures \ref{fig: pG straight string profile} and \ref{fig: pE straight string profile} respectively. Figure \ref{fig: pG eigenvector} represents an unstable perturbation to the string, while Figure \ref{fig: pE eigenvector} represents a perturbation that is stable.}
    \label{fig: eigenvectors}
\end{figure}

In Figure \ref{fig: eigenvectors} we show some examples of eigenvector solutions and we give the frequency and wavenumber that produces the correct associated eigenvalue. Figure \ref{fig: pG eigenvector} corresponds to an unstable perturbation of the string, while Figure \ref{fig: pE eigenvector} is a perturbation that is stable. An important point is that, in both cases, $\delta\hat{\sigma}$ is not the only non-negligible component of the eigenvector and it does not have the same radial profile as $|\sigma|$ (as seen in Figure \ref{fig:straight string profiles}) which means that they are not compatible with the assumptions made in \cite{Lemperiere2003a}.

In principle the stability of superconducting strings could be predicted using the following procedure:
\begin{itemize}
    \item The straight string solutions are calculated by numerically solving equations (\ref{eq: 1D EoMs start} - \ref{eq: 1D EoMs end}) for a particular parameter set and choice of $\chi$ (or $q_p$ to access the higher charge branch of solutions in the electric regime).
    \item A value for $\xi$ is chosen and a matrix is constructed, using the straight string solutions, which approximates the left hand side of equations (\ref{eq: delta phi equation} - \ref{eq: delta A equation}) when it acts upon the vector formed by the concatenation of $\delta\hat{\phi}$, $\delta\hat{\sigma}_1$, $\delta\hat{\sigma}_2$ and $\delta\hat{A}$.
    \item The eigenvalues and eigenvectors of this matrix can be calculated numerically, and the frequency $\nu$ and wavenumber $p$ can be determined using equation (\ref{eq: nu and p}). Solutions are unstable if either $\nu$ or $p$ is complex, but are only physically relevant instabilities if $\nu_2^2-p_2^2>0$. This process is repeated for other values of $\xi$ to scan the whole parameter space for unstable solutions.
\end{itemize}

In practice, we will want to carefully choose values for $\xi$ because a brute force approach will have to search through a two-dimensional parameter space, with unclear boundaries, and it is very computationally expensive to compute all of the eigenvalues for each value of $\xi$. We will usually be interested in looking at specific values of $p_1$ since that corresponds to different wavelengths of perturbation along the string. Unfortunately, in general, this does not directly translate to a particular value of $\xi$, due to the remaining freedom to vary $\nu_1$. This means that, for a particular choice of $\xi$, the eigenvalues need to be calculated before we will know the wavelength of the perturbation, which makes it difficult to know how much of the $\xi$ parameter space should be searched in order to test the stability of the string. 

Fortunately, by taking the complex conjugate of the differential equations, we can deduce the symmetry properties that when $\xi \to \xi^*$, $(\delta\hat{\phi},\delta\hat{\sigma}_1,\delta\hat{\sigma}_2,\delta\hat{A}_\theta,\Lambda) \to (\delta\hat{\phi}^*,\delta\hat{\sigma}_1^*,-\delta\hat{\sigma}_2^*,\delta\hat{A}_\theta^*,\Lambda^*)$ is a solution and similarly when $\xi \to -\xi^*$, $(\delta\hat{\phi},\delta\hat{\sigma}_1,\delta\hat{\sigma}_2,\delta\hat{A}_\theta,\Lambda) \to (\delta\hat{\phi}^*,\delta\hat{\sigma}_1^*,\delta\hat{\sigma}_2^*,\delta\hat{A}_\theta^*,\Lambda^*)$. Therefore, it is only necessary to consider one quadrant of the parameter space as the rest can be inferred by reflections about the axes. These symmetries also imply that the eigenvalues along the axes must come in complex conjugate pairs. 

In fact, by redefining $\delta\hat{A}_\theta = \sqrt{2}\rho \delta\hat{A}_\theta'$ and $\delta\hat{\sigma}_2 = i\delta\hat{\sigma}_2'$, the matrix formed by the discretisation of equations (\ref{eq: delta phi equation} - \ref{eq: delta A equation}) is very nearly symmetric, with the only non-symmetric parts being caused by the first order derivatives. It can be shown that, although this matrix is not symmetric, it is similar (in the mathematical sense where two matrices, $A$ and $B$, are similar if $B = P^{-1}AP$ for some change of basis matrix $P$) to a symmetric matrix. As similar matrices share the same eigenvalues and real, symmetric matrices have real eigenvalues, this means that the eigenvalues will all be real if $\xi$ is real. Note that the condition that $\nu_2<0$ for a solution to be exponentially growing rather than decaying is no longer important because one can flip the sign by a simple reflection.

This information is particularly useful for assessing the stability of strings in either the purely electric or purely magnetic frame. Although $\xi$, $\chi$ and $\Lambda$ are all invariant to Lorentz boosts along the string, these two frames are still useful because they have $\xi=-kp$ or $\xi=\omega\nu$, respectively. Half of the information about the mode is, therefore, known from the choice of $\xi$ and the other half can be worked out after the eigenvalues are calculated, partially solving the problem of searching for instabilities in $\xi$ space. In section \ref{sec: comparisons} we discuss in more detail how the purely magnetic or purely electric reference frames can be used to predict the boundary between stable and unstable eigenmodes.

A similar analysis was performed in \cite{Lemperiere2003a} for magnetic strings where they found that strings should be expected to be unstable, in the purely magnetic frame ($\omega=0$), if $k^2>\lambda_\sigma\Sigma_4/(4\Sigma_2)$. However, that analysis only allowed a restricted set of perturbations, $\delta\hat{\phi}=\delta \hat{A}_\theta=0$ and $\delta\hat{\sigma}_i=|\sigma|a_i$, where $i \in (1,2)$ and the $a_i$ are independent of $\rho$. We re-derive the results of \cite{Lemperiere2003a} from equations (\ref{eq: delta phi equation} - \ref{eq: delta A equation}) using these assumptions in appendix \ref{sec: prev analysis} and will be comparing the accuracy of our approach to \cite{Lemperiere2003a} in section \ref{sec: comparisons}. Note that the eigenmodes shown in Figure \ref{fig: eigenvectors} clearly violate these assumptions as there is a significant contribution from the vortex field and the radial profile of the perturbations to the condensate field is visually distinct from the underlying condensate, as seen in Figure \ref{fig:straight string profiles}.

\section{Simulations of pinching instabilities}\label{sec: sims}

In this section we will perform dynamical simulations of periodic straight superconducting strings to investigate how pinching instabilities develop and the ultimate effect that they have on the string. As previously stated this is equivalent to a vorton in many ways but removes the effects of curvature. We enforce axial symmetry and run two-dimensional evolution algorithms (and occasional full 3D simulations, to check that there is no difference with the 2D case), using equations (\ref{eq: phi1 eq} - \ref{eq: Atheta eq}), with periodic boundary conditions along the $z$ direction. Along the $\rho$ direction, we apply the boundary conditions $\phi_1(\rho=0)=0$, $\phi_1(\rho\to\infty)=\eta_\phi$, $\partial_\rho\sigma_{1,2}(\rho=0)=0$, $\sigma_{1,2}(\rho\to\infty)=0$, $A_\theta(\rho=0)=0$ and $A_\theta(\rho\to\infty)=n/g$. Note that the boundary conditions at infinity are enforced at a finite radius - $\rho_\text{max} = 40$ for parameter set G and $\rho_\text{max} = 100$ for parameter set E. We set the initial conditions to be the solutions to the static straight string equations with an applied perturbation using $\sigma_1 \to \sigma_1(1+\epsilon\cos(p_1z))$.

\begin{figure}[t]
    \centering
    \subfloat[$t=0$]{
        \centering
        \includegraphics[trim={4.5cm 3cm 4cm 7cm},clip,width=0.32\linewidth]{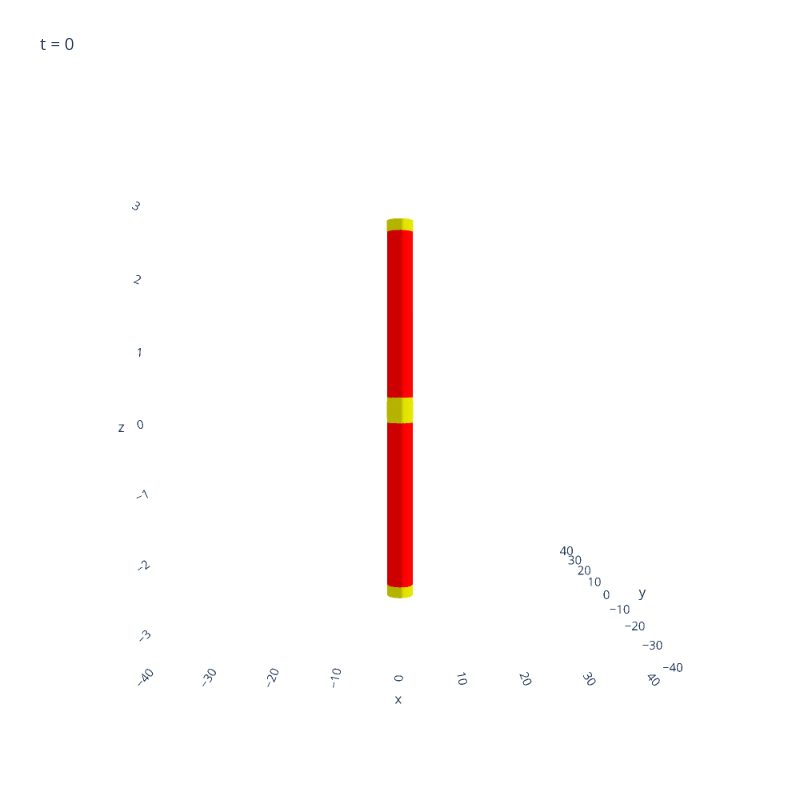}
    }
    \subfloat[$t=10.4$]{
        \centering
        \includegraphics[trim={4.5cm 3cm 4cm 7cm},clip,width=0.32\linewidth]{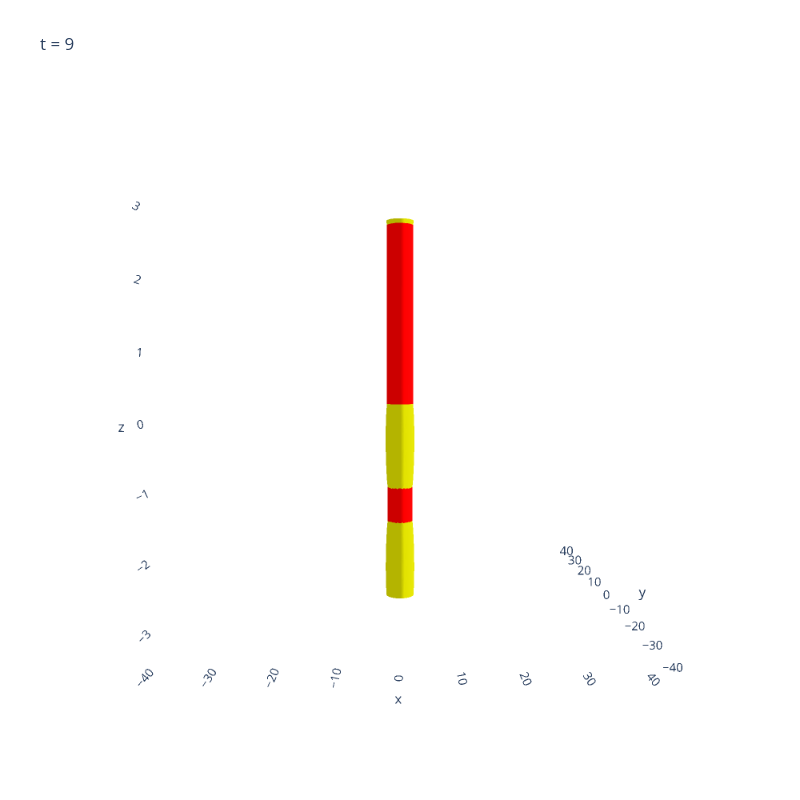}
    }
    \subfloat[$t=12$]{
        \centering
        \includegraphics[trim={4.5cm 3cm 4cm 7cm},clip,width=0.32\linewidth]{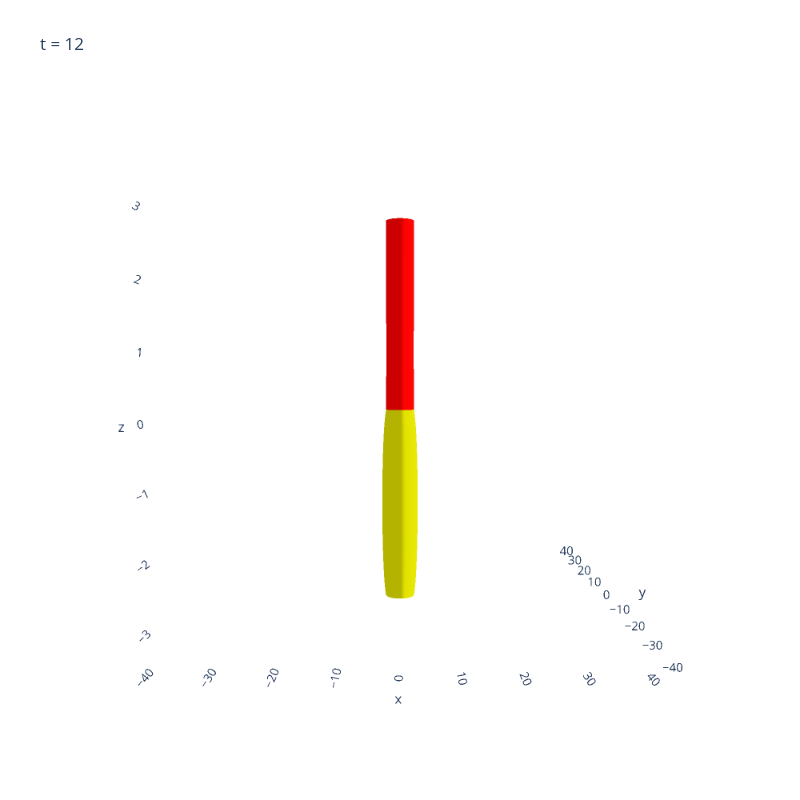}
    }
    \caption{Snapshots of a perturbed straight string, with periodic boundary conditions, that show the pinching instability growing and eventually causing the string to unwind. The mode being perturbed has a wavelength that is the same as the length of the string and an amplitude of $\epsilon = 0.01$. The red surface is an isosurface of $|\phi| = \frac{3}{5}$ while the yellow surface shows $\text{Re}(\sigma) = \frac{1}{5}\eta_\sigma$. Note that this is a relatively short section of string and that the $z$ axis has been elongated compared to $x$ and $y$ to better illustrate the evolution. The simulation was performed in the purely magnetic frame of the string shown in Figure \ref{fig: pG straight string profile} (parameter set G). The stability analysis predicts that the instability to this mode is driven by the eigenvector shown in Figure \ref{fig: pG eigenvector}.}
    \label{fig: pG isosurfaces}
\end{figure}

Using this approach, we have evolved the string shown in Figure \ref{fig: pG straight string profile}, in the purely magnetic frame, with an applied perturbation that has a wavelength equal to the length of the string and $\epsilon=0.01$. Figure \ref{fig: pG isosurfaces} shows isosurfaces of the fields during the simulation at $t=0$, $t=10.4$, by which point the instability has become clear, and at $t=12$ which shows the string after unwinding has occurred. The condensate initially had a winding number of $N=2$ (and no charge), which unwinds due to the instability and becomes $N=1$. This causes the string to go from the state with $\chi=-3$ to the less magnetic state with $\chi=-3/4$. Simulations of the $\chi=-3/4$ string suggest that this state is stable, but in this case the string unwinds a second time, shortly after the first, due to the significant deviations from the static solution created by the first unwinding event. It is unclear to us how generic this cascade of unwinding processes is (it may be that some strings do simply reduce their winding number and subsequently stabilise), but it would provide an efficient mechanism for unstable magnetic strings to become chiral, or at least significantly closer to chiral, while skipping over intermediate, stable magnetic strings.

\begin{figure}[t]
    \centering
    \subfloat[$t=0$]{
        \centering
        \includegraphics[trim={4.5cm 3cm 4cm 7cm},clip,width=0.32\linewidth]{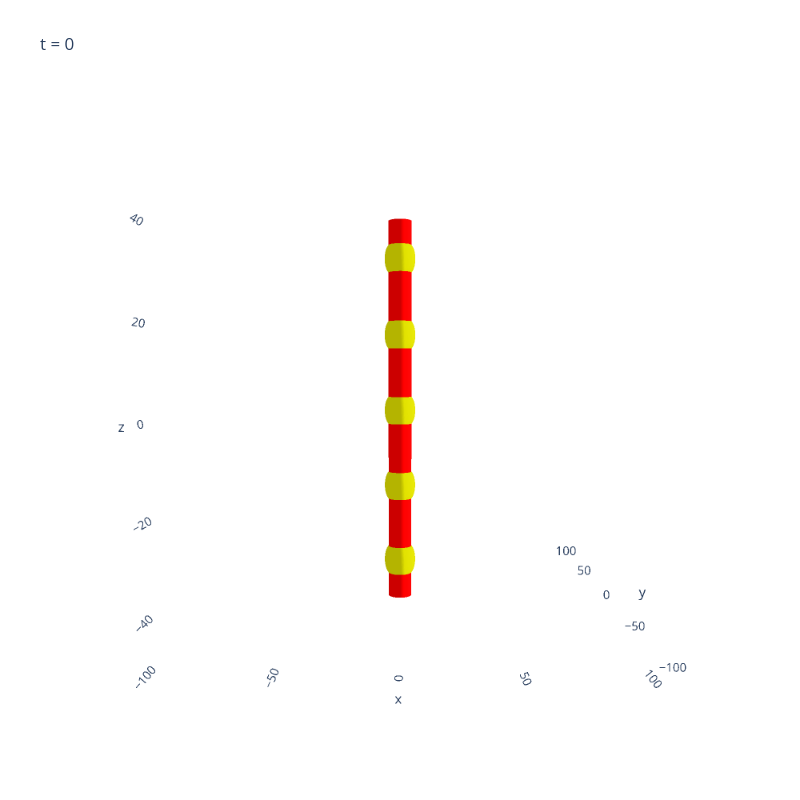}
    }
    \subfloat[$t=630$]{
        \centering
        \includegraphics[trim={4.5cm 3cm 4cm 7cm},clip,width=0.32\linewidth]{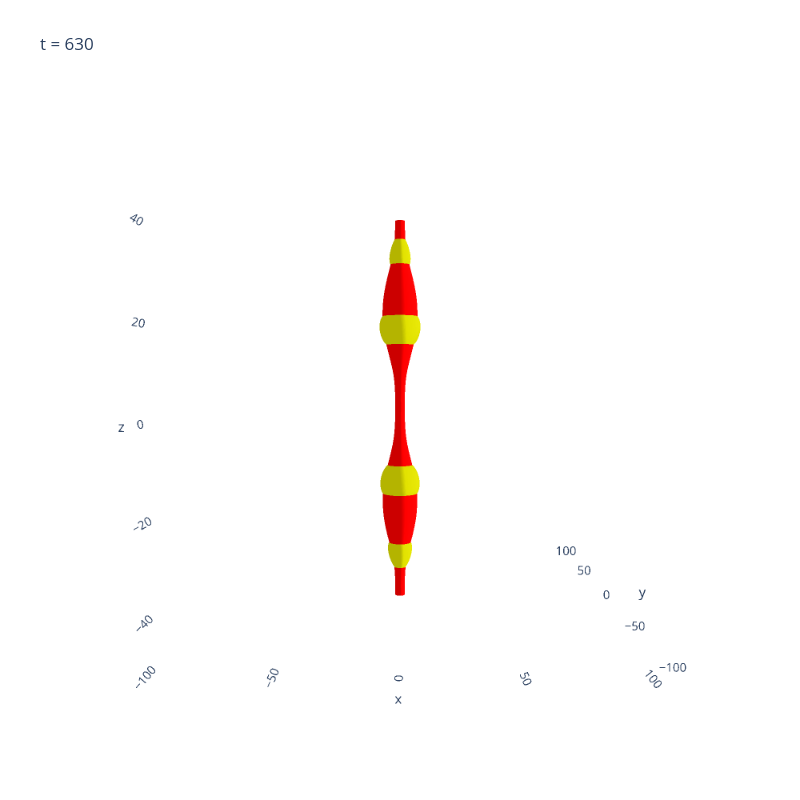}
        \label{fig: distorted}
    }
    \subfloat[$t=900$]{
        \centering
        \includegraphics[trim={4.5cm 3cm 4cm 7cm},clip,width=0.32\linewidth]{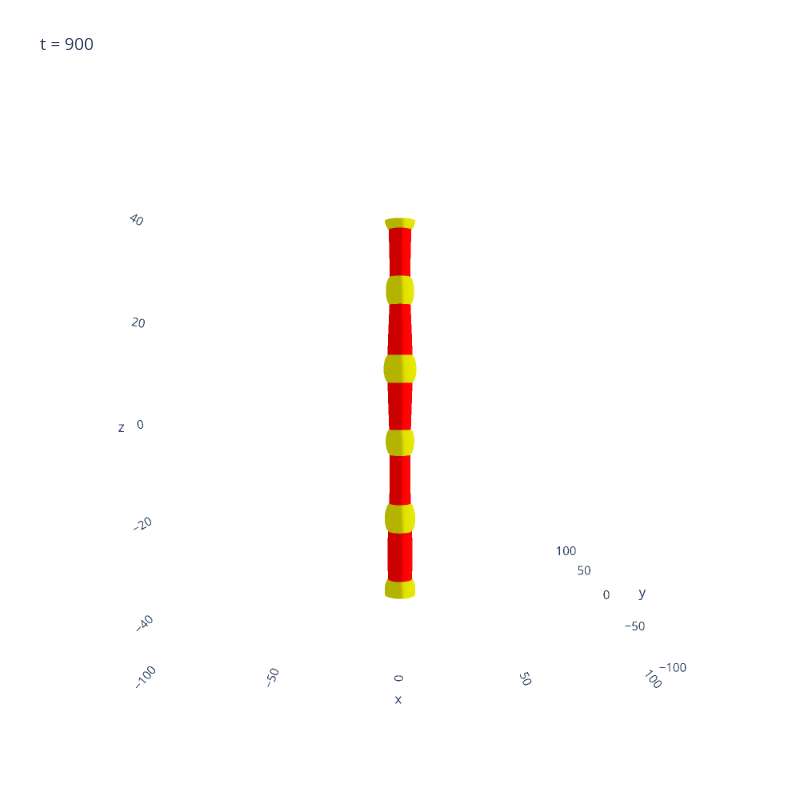}
    }
    \caption{Snapshots of a perturbed, electric, straight string (the one shown in Figure \ref{fig: pE straight string profile}) with periodic boundary conditions that show the pinching instability growing and causing the string to become highly distorted, although it later returns to a similar state to the one it started in. The instability is first evident at $t\sim 400$ and is at its most distorted by $t\sim 630$, as shown in the central plot. The perturbed mode has a wavelength that is half the length of the string and an amplitude of $\epsilon = 0.01$. The red surface is an isosurface of $|\phi| = \frac{3}{5}$ while the yellow surface shows $\text{Re}(\sigma) = \frac{1}{5}\eta_\sigma$.}
    \label{fig: pE isosurfaces}
\end{figure}

We have also evolved the electric string shown in Figure \ref{fig: pE straight string profile} after applying a perturbation with a wavelength that is half the length of the string and $\epsilon=0.01$, observing qualitatively different behaviour to the magnetic regime. In this case, we have not performed the simulation in the purely electric frame, instead we have used the frame in which the charge per unit length is $q=43$, $k=0.3$ and a length of string such that the winding number is $N=5$, so that the results can be directly compared with vorton simulations that will be discussed in section \ref{sec: vortons}. In Figure \ref{fig: pE isosurfaces}, we present snapshots of the isosurfaces at $t=0$, $t=630$ and $t=900$ that clearly display that the string is unstable, albeit in a very different way to the magnetic string.

These two simulations illustrate the qualitative differences between pinching instabilities in the magnetic and electric regimes. The electric string exhibits distinctive changes in width as the unstable mode grows, leading to a highly distorted shape in Figure \ref{fig: distorted} and implying that the vortex field must play a large role in the dynamics - which is not as obviously the case in the magnetic regime. However, the most interesting difference is that, although there is a clear instability in the electric string, it does not seem to ultimately lead to either unwinding events or the emission of charge - whereas in the magnetic regime it seems plausible that there is a mechanism for the string to move toward a chiral state. How instabilities in electric strings behave in the long term - for example whether it will continuously oscillate, slowly emit charge or relax into a new stationary state - is an interesting question, but not one that we will address in detail in this work. We focus instead on testing the accuracy of our predictions for the onset of instabilities, which we believe to be the more important issue.

\section{Comparing the stability analysis to simulations} \label{sec: comparisons}

We can test the accuracy of the stability analysis developed in section \ref{sec: gen analysis} by making comparisons between its predictions and simulations. The most important prediction to test is whether a given string is ultimately stable (in other words, stable to all wavelengths of perturbations) or not, but we can also make predictions for the critical wavenumber that represents the boundary between stability and instability, and also for the growth rate of unstable modes. In the magnetic regime, we will compare our simulations to predictions based on the analysis performed in \cite{Lemperiere2003a}. We will also make predictions, using the TSA, for the onset of longitudinal instabilities, and ultimately make the claim that they are the same as the pinching instabilities. 

We are particularly interested in investigating the parameter set $\eta_\sigma=0.61$, $\lambda_\sigma=10$, $\beta=3$ and $G=0.5$, which we named parameter set B in \cite{2022JHEP...04..005B} and constructed a close to chiral vorton solution that we have claimed was fully stable \cite{PhysRevLett.127.241601}. Pinching instabilities were present for small radii vortons, but they disappeared at larger radii. Therefore, determining whether chiral straight strings have pinching instabilities in this parameter set has implications for the stability of the vorton solution.

As our simulations have periodic boundary conditions, there is an additional constraint imposed upon the stability analysis that $p$ must be real (since $e^{-p_2z}$ is not periodic). As such, it is easiest to use our analysis to make predictions in the frame where $\omega = 0$, as it is then trivial to force $p$ to be real by simply only investigating real values of $\xi$. As previously explained, this has the additional benefit that the eigenvalues will all be real, so unstable perturbations simply correspond to $\Lambda < -p_1^2$ along this axis. The critical wavenumber is, therefore, given by the point along the $\xi$ axis where $-p_1^2$ is equal to the smallest possible eigenvalue.

\begin{figure}[!t]
    \centering
    \subfloat[The magnetic regime of the parameter set $\eta_\sigma=0.61$, $\lambda_\sigma=10$, $\beta=3$ and $G=0.5$ (parameter set B).]{
        \centering
        \includegraphics[trim={0.5cm 0 1.5cm 0},clip,width=0.46\linewidth]{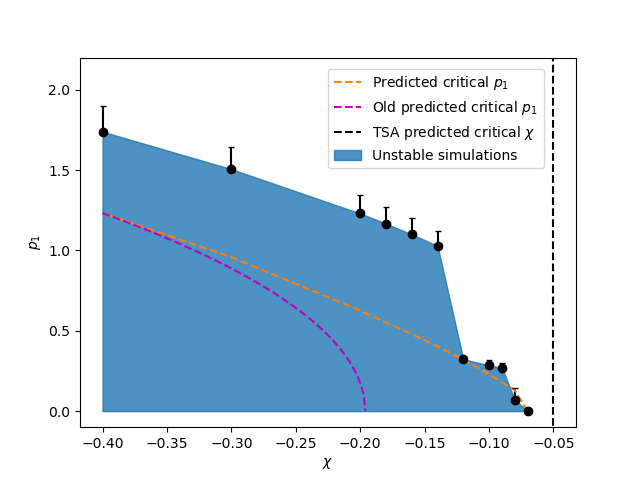}
        \label{fig: pB instability zones}
    }\hspace{1em}
    \subfloat[The electric regime of the parameter set $\eta_\sigma=1$, $\lambda_\sigma=2/3$, $\beta=2/3$ and $G=0.1$ (parameter set E).]{
        \centering
        \includegraphics[trim={0.5cm 0 1.5cm 0},clip,width=0.46\linewidth]{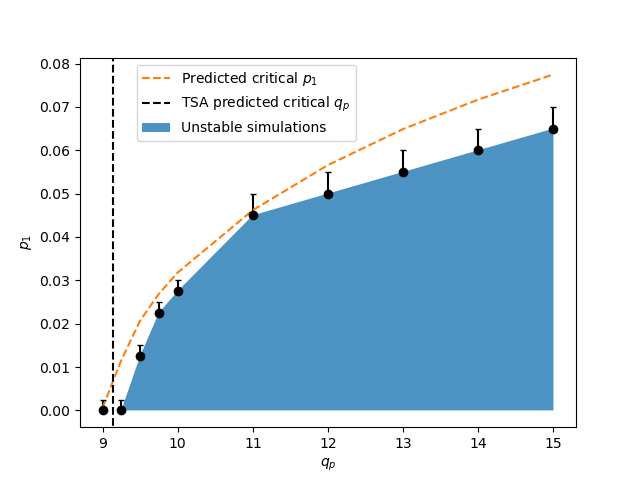}
        \label{fig: pE instability zones}
    }
    \caption{These plots compare the region of instability from both simulations and the predictions of various methods. The blue region shows the unstable region found by simulation - the top of the error bars indicate simulations in which the string was determined to be stable -  and the orange dotted line is the prediction of the critical value of $p_1$ from the method outlined in section \ref{sec: gen analysis}, below which there are expected to be instabilities. The magenta dotted line is an equivalent prediction, but using the method in \cite{Lemperiere2003a} - note that it only applies in the magnetic regime so it is only shown in figure \ref{fig: pB instability zones}. The black dotted line shows the prediction of a critical $\chi$ or $q_p$, with smaller values of $\chi$ or larger values of $q_p$ predicted to be unstable to longitudinal perturbations by the TSA. A mode is deemed to be ``stable" in our simulations if $|\sigma(\rho=0)|^2$ remains above $\frac{1}{10}\eta_\sigma^2$ by the end of the simulation. Simulations of magnetic strings are run until $t=1000$ and $t=10^4$ for electric strings, because those simulations are less computationally expensive.}
    \label{fig: instability zones}
\end{figure}

In Figure \ref{fig: pB instability zones} we show the regions of the parameter space for which the condensates unwind by $t=1000$ in our simulations of magnetic strings, after initially perturbing the mode with wavenumber $p_1$, and we compare this to the analytic prediction made in \cite{Lemperiere2003a}, our generalised analysis described in section \ref{sec: gen analysis} and the expectation from the TSA. Our simulations suggest that the strings are unstable to much larger values of $p_1$ than would be expected from our analysis, although the range of $\chi$ for which there are unstable strings is well predicted and is also more important because it determines the overall stability of the string. The sudden drop off of the blue region at high $\chi$ is a strange and unexpected feature that may indicate an issue with the simulations, or our method of determining whether the string is unstable. In particular, our criteria for detecting an instability - when $|\sigma|^2$ drops below $\frac{1}{10}\eta_\sigma^2$ at the centre of the string - probably doesn't work as well for highly magnetic strings since the magnitude of the condensate is smaller at the core. The onset of longitudinal instabilities predicted by the TSA is close to what we see in the simulations, and also the predictions from our stability analysis. The previous analysis, on the other hand, typically overestimates the range of $\chi$ for which there are stable strings.

Transforming to the frame in which $\omega=0$ is only possible in the magnetic regime. In the electric regime, we can transform instead to the frame in which $k=0$, which makes simulations easier, but the stability analysis harder. The simulations are easier because only one wavelength of the perturbation needs to be simulated, which is not the case for magnetic strings where the length of the string must be a common multiple of the wavelength associated with the winding and the wavelength of the perturbation, and the resolution must be sufficient to resolve the smaller of the two. Simulations of electric strings are, therefore, less computationally expensive, which allows us to run them over longer timescales. In particular, it is significantly easier to run simulations with long wavelength perturbations, allowing for more stringent testing of the TSA in precisely the regime in which it should be expected to work well.

Unfortunately, it is slightly more complicated to make predictions from our analysis in the electric case, as the wavelength of the mode is not known until after the eigenvalues are calculated, since only $\nu$ is specified by choosing a value for $\xi$. Additionally, the constraint that $p_2 = 0$ must still be satisfied due to the periodic boundaries. However, we know that real values of $\xi$ correspond to real eigenvalues, so we can find all modes with $\nu_2=p_2=0$ simply by searching along the real $\xi$ axis. A sensible criteria for the critical wavenumber would be when $-p_1^2$ is equal to the smallest possible eigenvalue, as in the magnetic case and this can only happen at $\xi=0$ in the purely electric frame. Although we have not ruled out the possibility for there to be unstable modes with larger $p_1$ elsewhere in the parameter space, we have not found any examples where this is the case, so we choose to define the critical wavenumber using the smallest eigenvalue at $\xi=0$.

Figure \ref{fig: pE instability zones} shows a comparison between the predictions of our analysis and the results of simulations of electric strings (the method used in \cite{Lemperiere2003a} is not shown here as it only applies to magnetic strings). We detect instabilities in the simulations if the magnitude of the condensate falls below some specified value (here we have used $|\sigma|^2<\frac{1}{10}\eta_\sigma^2$) before the end of the simulation at $t=10^4$. The TSA predicts that all strings to the right of the black dotted line should be unstable and our analysis predicts that strings are unstable to perturbations with wavenumbers below the orange dotted line. The two methods are in good agreement with each other and with the simulations. In particular, all three agree well, with only minor discrepancies, on the critical value of $q_p$, the charge per unit length in the purely electric frame, which is the most important feature as it determines whether each string will ultimately be stable or unstable.

Although there is moderate disagreement between the simulations and our analysis about the critical wavenumber, the crucial point is that, for both of the cases we have presented here, the stability analysis, simulations and prediction from the TSA are all in good agreement about which strings will ultimately be stable or unstable - corresponding to the intersection of the curves in Figure \ref{fig: instability zones} with the horizontal axis. This supports our claim that the pinching instabilities and longitudinal instabilities expected from the TSA are related. The TSA predicts that strings close to the chiral limit will not have longitudinal instabilities and our simulations and stability analysis also support this prediction for the parameter sets that we have tested. Vortons constructed from strings that are almost chiral, and are large enough for the effects of curvature to be negligible, should therefore also be stable to longitudinal perturbations.

From the stability analysis, we can additionally predict the growth rate of unstable modes, which can be compared to the simulations by estimating how long it will take for the perturbation to grow large enough to cause the condensate to unwind. We will only make this comparison for magnetic strings in parameter set B, as it is much harder to get a prediction for the growth rate in the electric regime. The reason for that is that in the magnetic regime, we can simply solve the eigenvalue problem of equations (\ref{eq: delta phi equation} - \ref{eq: delta A equation}) with $\xi = -kp_1$, and predict the growth rate in the purely magnetic frame from the most negative eigenvalue. In the electric regime however, we set $\xi = \omega\nu$ and then need to look for eigenvalues that satisfy $\Lambda = \nu^2 - p^2$, with $p_2=0$ and $p_1$ given by the mode that we will be perturbing in simulations (performed in the purely electric frame). In order to get a prediction for the growth rate, this process must be repeated with different choices of $\xi$ until we find the largest $\nu_2$ for which this is possible.

In order to make the comparison as simple as possible, we perturb the string with the normalised eigenvector solutions, which are multiplied by $\epsilon$ to control the magnitude of the perturbation. It is difficult to accurately predict when the condensate will unwind because non-linear effects kick in once the perturbation grows large enough. Nevertheless, we can make a simplistic estimate by setting $\sigma_2=0$ and calculating the time at which $\delta\sigma_1$ grows large enough that it reaches a fraction, $\alpha$, of the size of $|\sigma|$ at the core of the string. The requirement that $\sigma_2(\rho=0)=0$ implies that
\begin{equation}
    \delta\hat{\sigma}_1(\rho=0)\cos(p_1z + \nu_1t) - \delta\hat{\sigma}_2(\rho=0)\sin(p_1z + \nu_1t) = 0 ,
\end{equation}
which gives
\begin{equation}
    \varphi = \nu_1t + p_1z = \tan^{-1}\bigg(\frac{\text{Re}[\delta\hat{\sigma}_2(\rho=0)]}{\text{Im}[\delta\hat{\sigma}_2(\rho=0)]}\bigg) .
\end{equation}
The perturbation to $\sigma_1$ will then be equal to $\alpha|\sigma|$ at $\rho=0$ when
\begin{equation}
    t_c = \frac{\ln\epsilon}{\nu_2}-\frac{\ln\alpha}{\nu_2}-\frac{1}{\nu_2}\ln\bigg(\frac{|\sigma|(\rho=0)}{2(\text{Re}[\delta\hat{\sigma}_1(\rho=0)]\cos\varphi - \text{Im}[\delta\hat{\sigma}_1(\rho=0)]\sin\varphi)}\bigg), \label{eq: unwinding time}
\end{equation}
which we take as our prediction of the unwinding time. Although this is quite a crude estimate, with a parameter, $\alpha$, that must be fit to the simulation data, it allows the prediction of the growth rate to be well tested by examining the gradient of $t_c$ as a function of $\ln\epsilon$.

\begin{figure}[t]
    \centering
    \subfloat[Time taken until the condensate unwinds as a function of $\epsilon$, for $p_1=k/2$.]{
        \centering
        \includegraphics[trim={0.5cm 0 1.5cm 0},clip,width=0.46\linewidth]{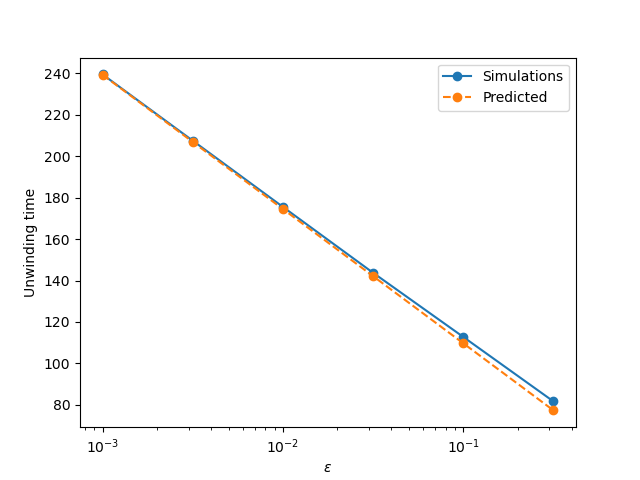}
        \label{fig: unwinding time vs epsilon}
    }\hspace{1em}
    \subfloat[The inverse of the unwinding time, $t_c^{-1}$ as a function of $p_1$.]{
        \centering
        \includegraphics[trim={0.1cm 0 1.5cm 0},clip,width=0.46\linewidth]{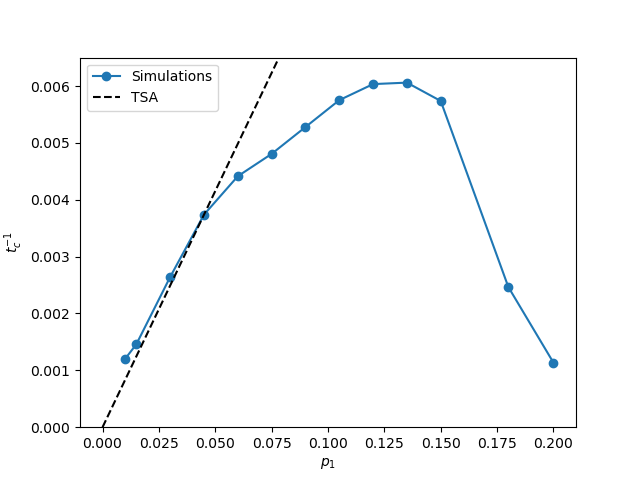}
        \label{fig: pB growth rates}
    }
    \caption{ These plots display how the time taken until the condensate unwinds changes with the amplitude of the perturbation and the wavenumber of the perturbations, respectively, for a string in parameter set B with $\chi=-0.09$ in the purely magnetic frame. This is marginally into the regime expected to be unstable. Figure \ref{fig: unwinding time vs epsilon} compares the predicted unwinding time from equation (\ref{eq: unwinding time}) with $\alpha=0.2$, to the unwinding time determined by simulations. The value of $\alpha$ may be chosen to match the data, since it relates to the specific unwinding criteria chosen, but the gradient of the line is a direct prediction from the stability analysis and clearly agrees well with the simulations. Figure \ref{fig: pB growth rates} shows how the inverse of the unwinding time (approximately proportional to the growth rate) changes with $p_1$. The growth rate grows roughly linearly with $p_1$ when it is small, as predicted by the TSA. The linear relationship predicted by the TSA is exemplified by the black dotted line, although it should be noted that this is for illustrative purposes only and is not calculated directly from the longitudinal sound speed due to the uncertainty in predicting the unwinding time from the growth rate. Smaller wavelength modes, where the TSA should not be expected to work well, break away from this trend and appear to be generally more stable than expected.}
    \label{fig: unwinding time plots}
\end{figure}

We present a plot of the unwinding time as a function of $\epsilon$ in Figure \ref{fig: unwinding time vs epsilon}, both from the theoretical predictions with $\alpha = 0.2$ and from the time taken until the condensate unwinds during simulations, for a string in parameter set B with $\chi=-0.09$. The simulations are performed by perturbing the straight string solutions with the eigenvectors that were calculated by the stability analysis, for the most unstable (largest $\nu_2$) solution with $p_1=k/2$, and the unwinding time was measured by detecting when $|\sigma|^2<\eta_\sigma^2/10$ somewhere along the string core. The value of $\alpha$ has simply been chosen to visually match the data, but it does take a reasonable value as it is large enough that second-order effects cannot be ignored. More importantly, the gradients of the lines are in very good agreement - with the simulations suggesting that $\nu_2 \approx -0.0364$ and the stability analysis predicting that $\nu_2 = -0.0355$.

In Figure \ref{fig: pB growth rates} we show how the time taken until the string unwinds varies with the wavenumber of the perturbation applied at $\chi = -0.09$, in parameter set B. We plot the inverse of the time taken because this will be roughly proportional to the growth rate of the mode and makes it clear that the expectation from the TSA that the growth rate should grow linearly with $p_1$ when it is small, is broadly correct. However, as the wavelength of the perturbation becomes comparable to the length scale of the string width, the growth rate breaks away from linear growth. In all of the simulations we have performed this has a stabilising effect on the string and our stability analysis predicts the same effect for all of the string solutions that we have checked.

\section{Vortons} \label{sec: vortons}

The results from our analysis should naturally apply to vortons as long as the radius is large enough that curvature effects can be ignored. We have already discussed that this seems to be true for the vorton we presented in \cite{PhysRevLett.127.241601}, which does not exhibit pinching instabilities if the vorton is large enough that the effects of curvature are negligible. We have also previously presented an electric vorton solution \cite{2022JHEP...04..005B}, constructed from a string on the higher charge branch, that the TSA predicts should be unstable to both extrinsic and pinching instabilities. In this case, simulations of the corresponding straight string did show signs of pinching instabilities - this is the simulation shown in Figure \ref{fig: pE isosurfaces}. In contrast to the previous example, we should expect to see these pinching instabilities in simulations regardless of the size of the vorton.

Unfortunately, as this vorton is also unstable to extrinsic perturbations, this can interfere by destroying the vorton before the pinching instability becomes evident. There are two ways to make the effect more clear. Either perturb the vorton with a mode to which it is not extrinsically unstable, which may not always be practical depending upon the vorton in question, or use the fact that the growth rate of the extrinsic instability is inversely proportional to the vorton radius, while the growth rate of the pinching instability should not change (although probably will, to a small degree, due to curvature effects). In this case, from the simulations presented in Figure \ref{fig: pE isosurfaces}, we should expect that the pinching instability will be evident by $t\sim 400$ and show the most dramatic effects by $t\sim 600$.

\begin{figure}[t!]
    \centering
    \subfloat[$t=0$]{
        \centering
        \includegraphics[trim={4.5cm 3cm 4cm 7cm},clip,width=0.32\linewidth]{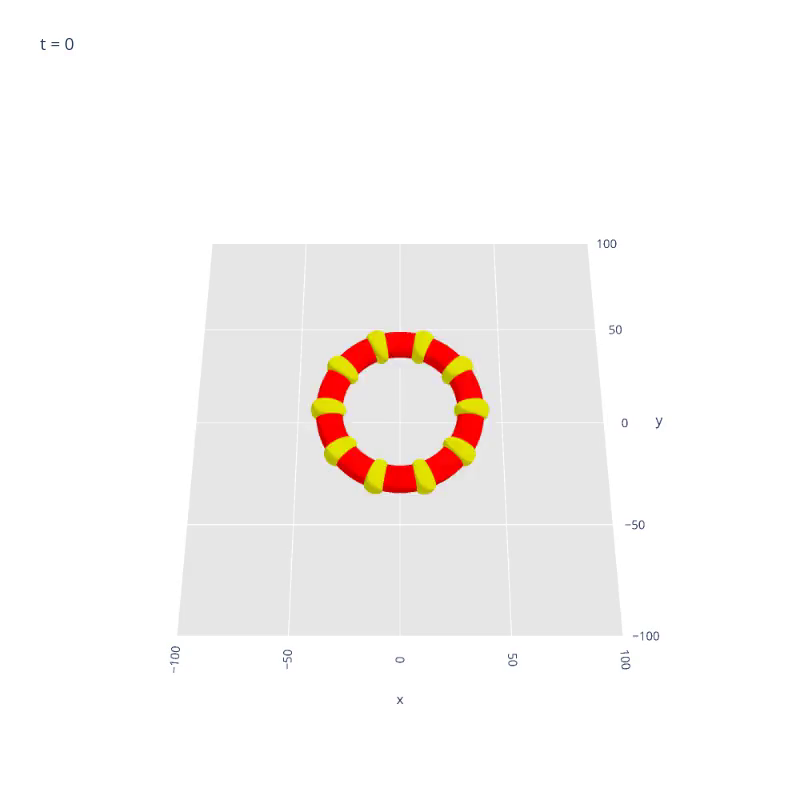}
    }\hfill
    \subfloat[$t=300$]{
        \centering
        \includegraphics[trim={4.5cm 3cm 4cm 7cm},clip,width=0.32\linewidth]{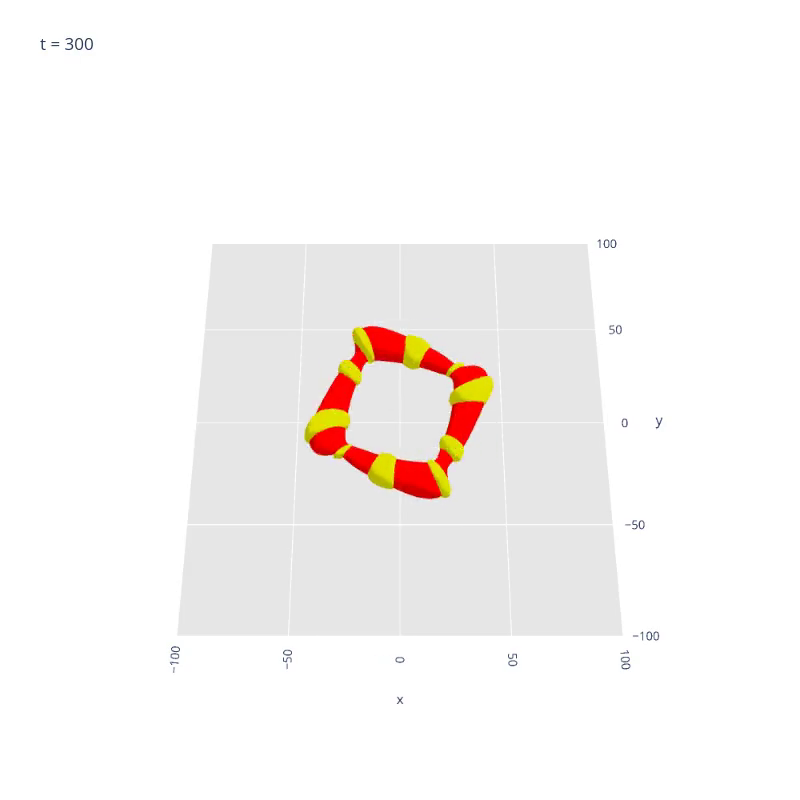}
    }\hfill
    \subfloat[$t=460$]{
        \centering
        \includegraphics[trim={4.5cm 3cm 4cm 7cm},clip,width=0.32\linewidth]{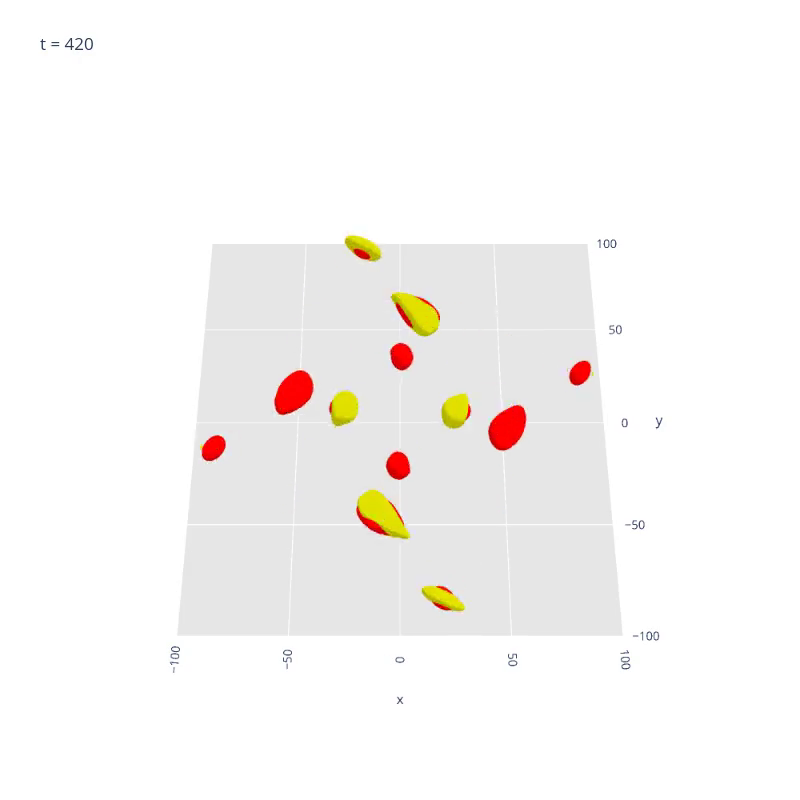}
    }\hfill\vspace{-0.4cm}
    \subfloat[$t=0$]{
        \centering
        \includegraphics[trim={4.5cm 3cm 4cm 7cm},clip,width=0.32\linewidth]{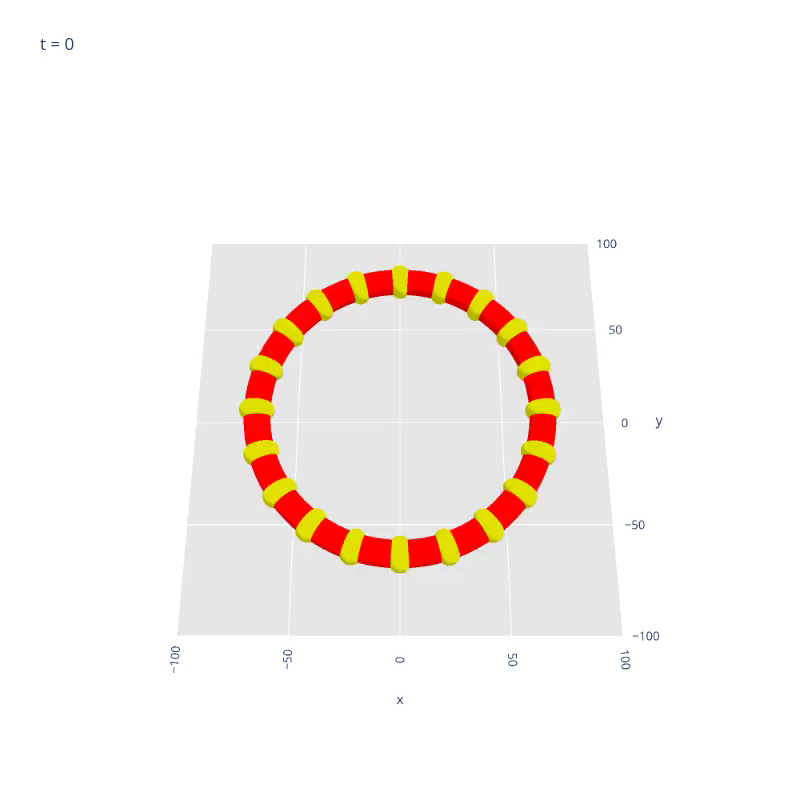}
    }\hfill
    \subfloat[$t=600$]{
        \centering
        \includegraphics[trim={4.5cm 3cm 4cm 7cm},clip,width=0.32\linewidth]{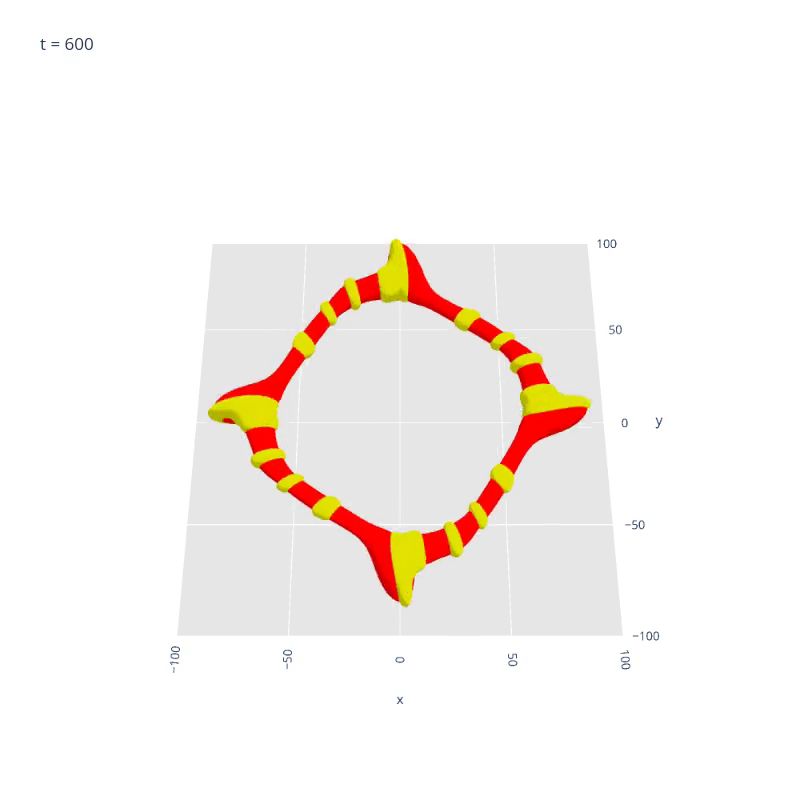}
    }\hfill
    \subfloat[$t=780$]{
        \centering
        \includegraphics[trim={4.5cm 3cm 4cm 7cm},clip,width=0.32\linewidth]{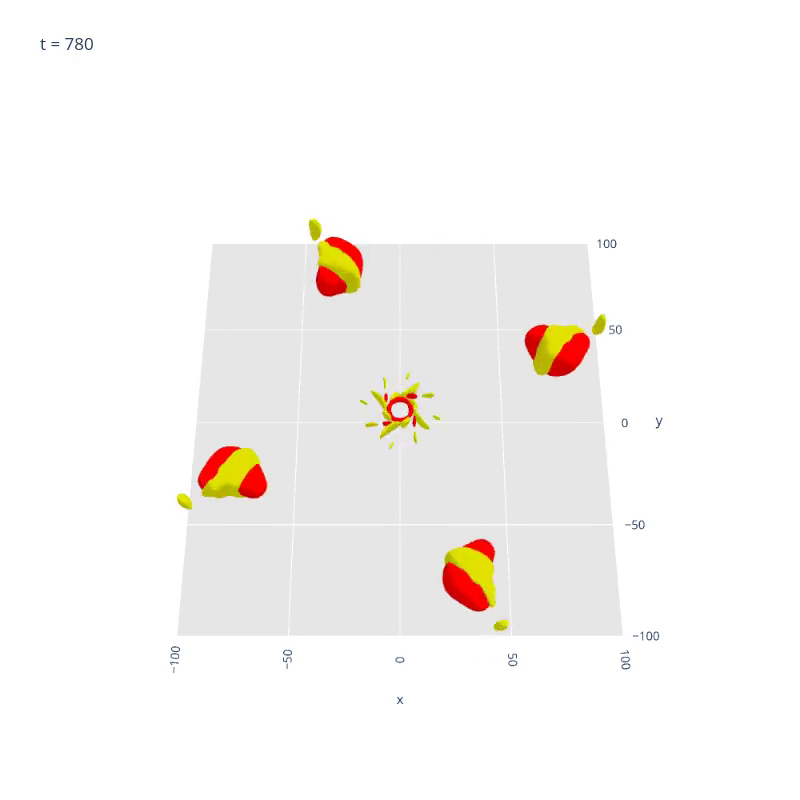}
    }\hfill\vspace{-0.4cm}
    \subfloat[$t=0$]{
        \centering
        \includegraphics[trim={4.5cm 3cm 4cm 7cm},clip,width=0.32\linewidth]{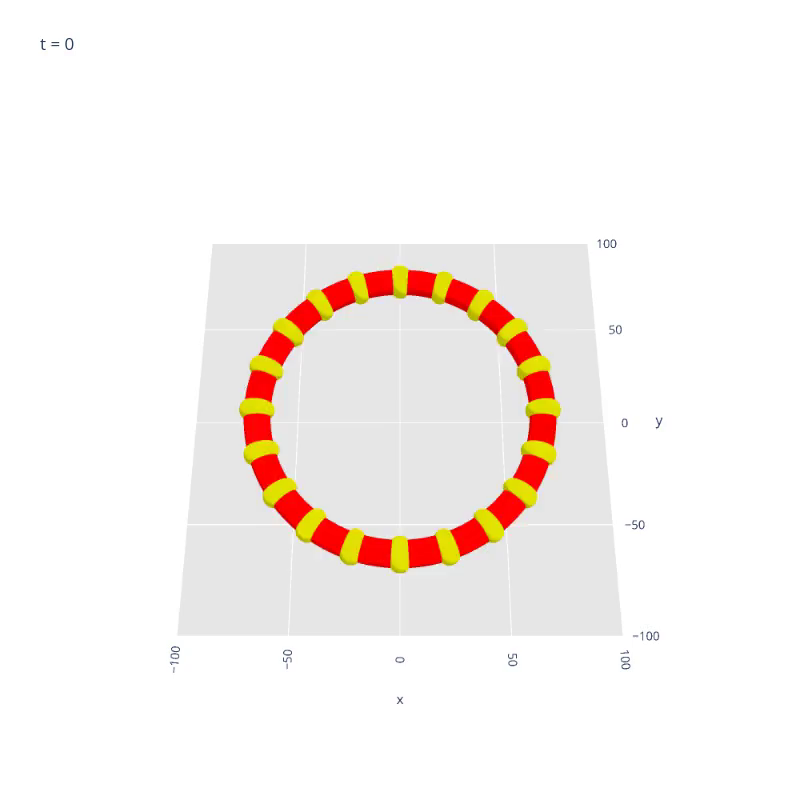}
    }\hfill
    \subfloat[$t=500$]{
        \centering
        \includegraphics[trim={4.5cm 3cm 4cm 7cm},clip,width=0.32\linewidth]{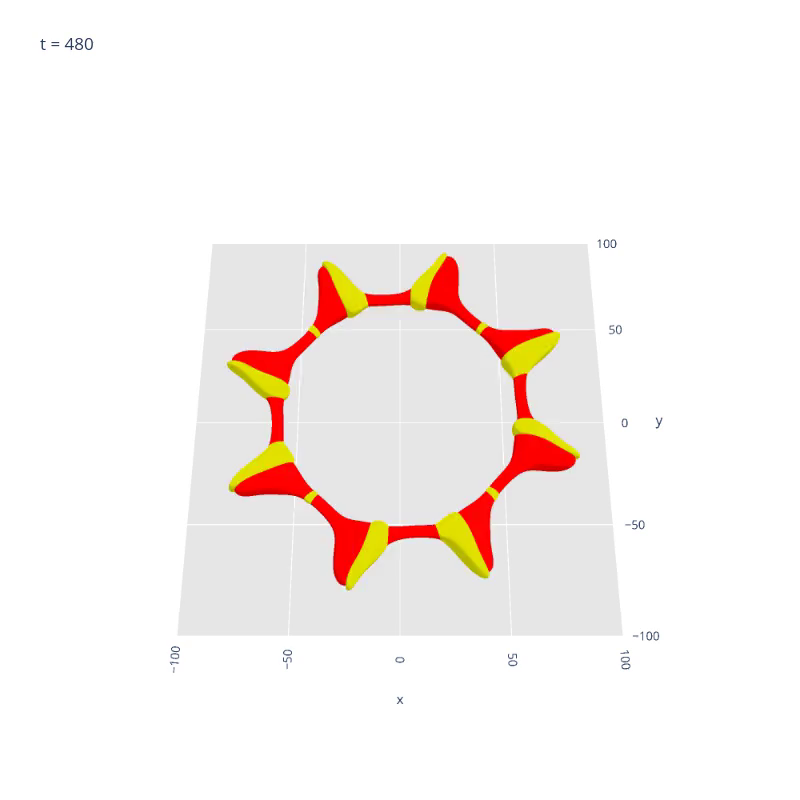}
    }\hfill
    \subfloat[$t=1000$]{
        \centering
        \includegraphics[trim={4.5cm 3cm 4cm 7cm},clip,width=0.32\linewidth]{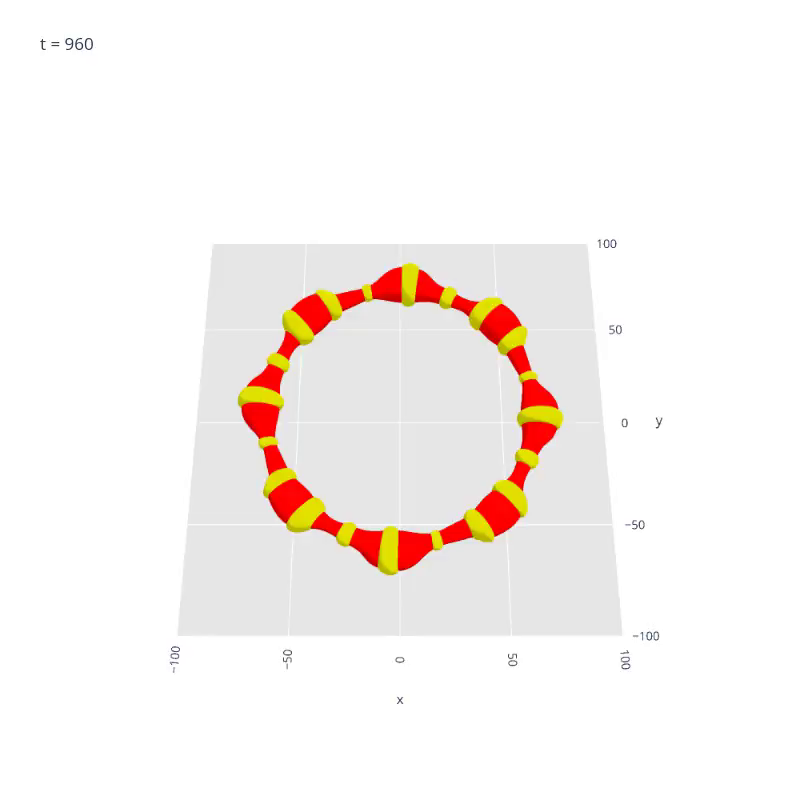}
    }\hfill
    \caption{Snapshots of vorton simulations that show the development of both extrinsic and pinching instabilities. All vortons shown have approximately the same $Q/L$ and $N/L$ and, therefore, the TSA predicts that they should have approximately the same stability properties. The top row shows a vorton with $Q=9000$ and $N=10$ that has an obvious extrinsic instability to the $m=4$ mode that completely destroys the vorton by $t=460$. There are some possible signs of pinching instabilities at $t=300$, but it is not completely clear and the extrinsic instability is the dominating effect. As we do not expect the timescale of the pinching instability to be affected by the size of the vorton, but the growth rate of the extrinsic instability is predicted to scale with $L^{-1}$, we can attempt to make the pinching instability dominate by simulating a larger vorton. To this end, both the middle and bottom rows show a larger vorton with $Q=18000$ and $N=20$, but the middle row shows a simulation where an $m=4$ mode (the same extrinsic instability as the top row) has been perturbed while an $m=8$ mode (the same intrinsic instability as the top row, since $p_1 = 2\pi m/L$) has been perturbed in the bottom row. The middle row snapshots do display clear effects of a pinching instability at $t=600$, but the vorton is still ultimately destroyed in a similar way to the top row. In contrast, the bottom row shows dramatic signs of a pinching instability at $t=500$ and, just like in our simulations of straight electric strings (see Figure \ref{fig: pE isosurfaces}), the vorton is not destroyed by these large distortions and returns to the calmer state seen at $t=1000$.}
    \label{fig: vorton snapshots}
\end{figure}

In Figure \ref{fig: vorton snapshots} we show isosurface snapshots from three different simulations of vortons that all have the same $Q/L$ and $N/L$ as the straight string from Figure \ref{fig: pE isosurfaces}. The top row shows a vorton with $Q=9000$ and $N=10$ while the middle and bottom rows show a vorton with $Q=18000$ and $N=20$. The top and middle rows both show simulations in which an $m=4$ mode has been excited, and therefore we should expect to see similar effects from the extrinsic instabilities (albeit with a longer time scale for the larger vorton) while the bottom row shows a vorton with an excited $m=8$ mode and should show similar effects from the pinching instabilities as the top row because $p_1 = 2\pi m/L$. The two upper rows both show some signs of pinching instabilities, but are ultimately dominated by the extrinsic instability which causes the vorton to be destroyed. However, the isosurfaces shown in the bottom row are very similar to what was seen in the straight string simulation, with the perturbations becoming clear by $t\sim 400$ and the distortions reaching their greatest by $t\sim 500$. The vorton is not destroyed by the end of our simulation but instead returns back to a less disrupted state - in agreement with the expectations from Figure \ref{fig: pE isosurfaces}.

\section{Conclusions}

We have presented a new analysis of pinching instabilities that takes the width of the string into account and predicts which wavelengths (if any) a given string will be unstable to. This appears to work very well in the electric regime, but underestimates the range of unstable $p_1$ in the magnetic regime. It does, however, make good predictions about which strings will ultimately be unstable, and also makes connections with longitudinal perturbations in the TSA, which is clearly more important.

There are still some remaining unanswered questions about the implications of pinching instabilities for the ultimate fate of the string. In the magnetic case, we observed a cascade of unwinding events which resulted in a strongly magnetic string ultimately becoming chiral, even though there were stable magnetic states in between. We have not established whether this was a special case or if it is a general feature of pinching instabilities. If it is general, it would allow chiral strings to be easily produced. On the other hand, electric strings with pinching instabilities do not seem to resolve the problem of having too much charge and we instead see them simply oscillate between the unexcited string and a highly distorted shape, rather than emitting charge and relaxing to a less electric string. There are many possibilities for the long term dynamics of these strings that our simulations are not sufficient to distinguish between, for example charge could be emitted very slowly from the string or the oscillation could indicate a new stationary state that looks like a Q-ball attached to a string (so that all the charge has clumped in one place). More simulations, particularly ones over longer time periods and for different parameter sets, will be required to better understand this issue and we leave this open for future work.

We emphasize that the most important result obtained in this paper is that the TSA makes reasonably accurate predictions about the onset of pinching instabilities (as long as the curvature can be neglected) and that small wavelength perturbations, where the TSA should not be relied upon due to the non-negligible width of the string, are typically stable. Although we cannot say for certain that this is always the case for different sets of parameters, it has been the case in all sets that we have checked and is certainly the case for the stable vorton solution, discovered in parameter set B, that we discussed in \cite{PhysRevLett.127.241601,2022JHEP...04..005B}. More generally, as the TSA predicts stability to both extrinsic and intrinsic types of perturbations to vortons in the chiral limit, we expect that if a given parameter set allows for chiral superconducting strings to exist, then it will contain stable vortons.

\FloatBarrier

\appendix

\section{Ansatz} \label{sec: ansatz}

Assuming nothing but axial symmetry in the fields, except for a phase winding of $n$ in the $\phi$ field, and using the temporal gauge so that $A_t=0$, we have the general ansatz $\phi = (\phi_1(t,\rho,z) + i\phi_2(t,\rho,z))e^{in\theta}$, $\sigma = (\sigma_1(t,\rho,z) + i\sigma_2(t,\rho,z))e^{i(\omega t + kz)}$ and $A_i = A_i(t,\rho, z)$, for which the equations of motion are

\begin{gather}
    \begin{split}
        \frac{\partial^2\phi_1}{\partial t^2} - \frac{\partial^2\phi_1}{\partial\rho^2} - \frac{1}{\rho}\frac{\partial\phi_1}{\partial\rho} - \frac{\partial^2\phi_1}{\partial z^2} -2g\bigg(A_\rho\frac{\partial\phi_2}{\partial\rho} + A_z\frac{\partial\phi_2}{\partial z}\bigg) - g\phi_2\bigg(\frac{\partial A_\rho}{\partial\rho} + \frac{A_\rho}{\rho} + \frac{\partial A_z}{\partial z}\bigg) \\ + \bigg[\frac{1}{2}\lambda_\phi(\phi_1^2 + \phi_2^2 - \eta_\phi^2) + \beta(\sigma_1^2 + \sigma_2^2) + \bigg(\frac{n - gA_\theta}{\rho}\bigg)^2 + g^2(A_\rho^2 + A_z^2)\bigg]\phi_1 = 0 ,
    \end{split} \\
    \begin{split}
        \frac{\partial^2\phi_2}{\partial t^2} - \frac{\partial^2\phi_2}{\partial\rho^2} - \frac{1}{\rho}\frac{\partial\phi_2}{\partial\rho} - \frac{\partial^2\phi_2}{\partial z^2} +2g\bigg(A_\rho\frac{\partial\phi_1}{\partial\rho} + A_z\frac{\partial\phi_1}{\partial z}\bigg) + g\phi_1\bigg(\frac{\partial A_\rho}{\partial\rho} + \frac{A_\rho}{\rho} + \frac{\partial A_z}{\partial z}\bigg) \\ + \bigg[\frac{1}{2}\lambda_\phi(\phi_1^2 + \phi_2^2 - \eta_\phi^2) + \beta(\sigma_1^2 + \sigma_2^2) + \bigg(\frac{n - gA_\theta}{\rho}\bigg)^2 + g^2(A_\rho^2 + A_z^2)\bigg]\phi_2 = 0 ,
    \end{split} \\
    \frac{\partial^2\sigma_1}{\partial t^2} - \frac{\partial^2\sigma_1}{\partial\rho^2} - \frac{1}{\rho}\frac{\partial\sigma_1}{\partial\rho} - \frac{\partial^2\sigma_1}{\partial z^2} - 2\omega\frac{\partial\sigma_2}{\partial t} + 2k\frac{\partial\sigma_2}{\partial z} + \bigg[\frac{1}{2}\lambda_\sigma(\sigma_1^2 + \sigma_2^2 - \eta_\sigma^2) + \beta(\phi_1^2 + \phi_2^2) - \omega^2 + k^2\bigg]\sigma_1 = 0, \\
    \frac{\partial^2\sigma_2}{\partial t^2} - \frac{\partial^2\sigma_2}{\partial\rho^2} - \frac{1}{\rho}\frac{\partial\sigma_2}{\partial\rho} - \frac{\partial^2\sigma_2}{\partial z^2} + 2\omega\frac{\partial\sigma_1}{\partial t} - 2k\frac{\partial\sigma_1}{\partial z} + \bigg[\frac{1}{2}\lambda_\sigma(\sigma_1^2 + \sigma_2^2 - \eta_\sigma^2) + \beta(\phi_1^2 + \phi_2^2) - \omega^2 + k^2\bigg]\sigma_2 = 0, \\
    \frac{\partial^2A_\rho}{\partial t^2} - \frac{\partial^2A_\rho}{\partial z^2} + \frac{\partial^2A_z}{\partial\rho\partial z} +2g\bigg(\phi_2\frac{\partial\phi_1}{\partial\rho} - \phi_1\frac{\partial\phi_2}{\partial\rho}\bigg) + 2g^2A_\rho(\phi_1^2 + \phi_2^2) = 0, \\
    \frac{\partial^2 A_\theta}{\partial t^2} - \frac{\partial^2A_\theta}{\partial\rho^2} + \frac{1}{\rho}\frac{\partial A_\theta}{\partial\rho} - \frac{\partial^2A_\theta}{\partial z^2} - 2g(n-gA_\theta)(\phi_1^2 + \phi_2^2) = 0, \\
    \frac{\partial^2A_z}{\partial t^2} - \frac{\partial^2A_z}{\partial\rho^2} - \frac{1}{\rho}\frac{\partial A_z}{\partial\rho} + \frac{\partial^2A_\rho}{\partial\rho\partial z} + \frac{1}{\rho}\frac{\partial A_\rho}{\partial z} + 2g\bigg(\phi_2\frac{\partial\phi_1}{\partial z} - \phi_1\frac{\partial\phi_2}{\partial z}\bigg) + 2g^2A_z(\phi_1^2 + \phi_2^2) = 0.
\end{gather}

By inspecting these equations, it becomes clear that $\phi_2 = A_\rho = A_z \equiv 0$ is a self-consistent solution and therefore perturbations to any of the other fields will not induce any changes in these field components. The straight string static solutions have no $t$ or $z$ dependence and so setting these derivatives to zero allows for $\sigma_2$ to also be consistently set to zero and the one dimensional equations of motion (\ref{eq: 1D EoMs start} - \ref{eq: 1D EoMs end}) are recovered. However, $\sigma_2$ will be perturbed away from zero if there are $t$ or $z$ dependent perturbations in $\sigma_1$, which are exactly the type that we would like to investigate. The equations of interest are therefore those given by equations (\ref{eq: phi1 eq} - \ref{eq: Atheta eq}). 

\section{Derivation of results in \cite{Lemperiere2003a}} \label{sec: prev analysis}

In order to derive the results of \cite{Lemperiere2003a} from our analysis, we can set $\delta\hat{\phi}=\delta\hat{A}_\theta = 0$, $\delta\hat{\sigma}_1 = |\sigma|a_1$ and $\delta\hat{\sigma}_2 = |\sigma|a_2$ where $a_1$ and $a_2$ are complex constants and $|\sigma|=|\sigma|(\rho)$ satisfies the straight string equations. The perturbation equations become
\begin{align}
    -\frac{d^2|\sigma|}{d\rho^2}a_1 - \frac{1}{\rho}\frac{d|\sigma|}{d\rho}a_1 + \bigg[\frac{1}{2}\lambda_\sigma(3|\sigma|^2 - \eta_\sigma^2) + \beta|\phi|^2 - \chi\bigg]|\sigma|a_1 - 2i\xi|\sigma|a_2 &= \Lambda|\sigma|a_1 , \label{eq: LS perts 1} \\
    -\frac{d^2|\sigma|}{d\rho^2}a_2 - \frac{1}{\rho}\frac{d|\sigma|}{d\rho}a_2 + \bigg[\frac{1}{2}\lambda_\sigma(|\sigma|^2 - \eta_\sigma^2) + \beta|\phi|^2 - \chi\bigg]|\sigma|a_2 + 2i\xi|\sigma|a_1 &= \Lambda|\sigma|a_2 , \label{eq: LS perts 2}
\end{align}
and now substituting equation (\ref{eq: sigma straight static EoM}) into these expressions, multiplying through by $|\sigma|$ and integrating over the radial direction yields the matrix equation, 
\begin{equation}
    \begin{pmatrix}
        \lambda_\sigma\Sigma_4 - \Lambda\Sigma_2 & -2i\xi\Sigma_2 \\
        2i\xi\Sigma_2 & -\Lambda\Sigma_2
    \end{pmatrix}
    \begin{pmatrix}
        a_1 \\ a_2
    \end{pmatrix}
    = 0 ,
\end{equation}
There are only non-trivial eigenvector solutions to this equation when the determinant is zero. For magnetic strings in the frame with $\omega=0$, this can be achieved by setting
\begin{equation}
    \nu^2 = \frac{\lambda_\sigma\Sigma_4}{2\Sigma_2} + p_1^2 \pm \sqrt{\bigg(\frac{\lambda_\sigma\Sigma_4}{2\Sigma_2}\bigg)^2 + 4k^2p_1^2} .
\end{equation}
Unstable modes have complex $\nu$ which leads to the instability condition
\begin{equation}
    p_1^2 < 4k^2 - \frac{\lambda_\sigma\Sigma_4}{\Sigma_2} ,
\end{equation}
and therefore, this analysis suggests that the pinching instability only exists when
\begin{equation} \label{eq: LS condition}
    k^2 > \frac{\lambda_\sigma\Sigma_4}{4\Sigma_2}.
\end{equation}
From our simulations (see section \ref{sec: sims}), we have found that this expression is not very accurate at predicting the onset of instabilities, although we did find it useful for ball-park estimates since it is very easy to calculate. The assumptions made are overly restrictive, which can be made very clear by examining equations (\ref{eq: LS perts 1} - \ref{eq: LS perts 2}) in more detail. The substitution of equation (\ref{eq: sigma straight static EoM}) results in the two equations,
\begin{align}
    [\lambda_\sigma|\sigma|^3 - \Lambda|\sigma|]a_1 - 2i\xi|\sigma|a_2 &= 0 , \\
    -\Lambda|\sigma|a_2 + 2i\xi|\sigma|a_1 &= 0 ,
\end{align}
that only have a solution if $|\sigma|(\rho)$ is a constant, which does not occur except in the trivial case of a non-superconducting string with $|\sigma|(\rho)=0$.

\bibliographystyle{ieeetr}
\bibliography{refs.bib}

\end{document}